\documentclass[aps,prb,reprint]{revtex4-1}

\usepackage{amsmath}    
\usepackage{times,mathptmx} 
\usepackage{graphicx}   
\usepackage{verbatim}   
\usepackage{color}      
\usepackage{subfigure}  
\usepackage{hyperref}   
\usepackage{CJK} 

\begin{document}
\begin{CJK*}{UTF8}{}

\title{Dirac fermion optics and directed emission from single- and bilayer graphene cavities}
\author{Jule-Katharina Schrepfer}
\affiliation{Institute for Physics and Institute for Micro- and Nanotechnologies, Technische Universit\"at Ilmenau, D-98693 Ilmenau, Germany}
\author{Szu-Chao Chen (\CJKfamily{bsmi}{陳思超})}
\affiliation{Department of Physics, National Cheng Kung University, Tainan 70101, Taiwan}
\author{Ming-Hao Liu (\CJKfamily{bsmi}{劉明豪})}
\affiliation{Department of Physics, National Cheng Kung University, Tainan 70101, Taiwan}
\author{Klaus Richter}
\affiliation{Institute of Theoretical Physics, University of Regensburg, 93040 Regensburg, Germany}
\author{Martina Hentschel}
\email{martina.hentschel@physik.tu-chemnitz.de}
\affiliation{Institute of Physics, 
Technische Universit\"at Chemnitz, 
D-09107 Chemnitz, Germany}

\begin{abstract}

High-mobility graphene hosting massless charge carriers with linear dispersion provides a promising platform for electron optics phenomena.   
Inspired by the physics of dielectric optical micro-cavities where the photon emission characteristics can be efficiently tuned via the cavity shape, we study corresponding mechanisms for trapped Dirac fermionic resonant states in deformed micro-disk graphene billiards and directed emission from those. 
In such graphene devices a back-gate voltage provides an additional tunable parameter to mimic different effective refractive indices and thereby the corresponding Fresnel laws at the boundaries. Moreover, cavities based on single-layer and double-layer graphene exhibit Klein- and
anti-Klein tunneling, respectively, leading to distinct differences with respect to dwell times and resulting emission profiles of the cavity states.
Moreover, we find a variety of different emission characteristics depending on the position of the source where charge carriers are fed into the cavites.
Combining quantum mechanical simulations with optical ray tracing and a corresponding phase-space analysis, we demonstrate strong confinement of the emitted charge carriers in the mid field of single-layer graphene systems and can relate this to a lensing effect. 
For bilayer graphene, trapping of the resonant states is more efficient and the emission characteristics do less depend on
the source position.
\end{abstract}

\maketitle
\end{CJK*}

\section{Introduction}

Due to its linear energy-momentum relation for low-energy excitations, graphene provides an ideal   low-dimensional condensed-matter platform for Dirac electron optics: Comparing graphene's  linear dispersion $E(k) = \pm  \hbar v_{\rm F} k$ with the Planck-Einstein
relation for photons in vacuum, $E(k) = \hbar ck$, optics-like electron physics in 
graphene is naturally expected. Here $c$ is the speed of light and $v_{\rm F}$ is the 
Fermi velocity of electrons in graphene. Hence, the electronic states in graphene carry certain features of photons but, at the same time, respond to external electric and magnetic fields. Furthermore, graphene electronics is similar to semiconductor physics
with respect to its carrier polarity. However, the gapless energy band structure of
graphene makes it much easier to switch between n- and p-type states by electrical
charging and depleting, enabling efficient gating of graphene hetero-junctions. As a result, the combination of these various special properties, i.e.\ that charge carriers
in graphene partly behave like photons, are deflected by magnetic fields, are reflected or diffracted at p-n junctions and propagate dispersionless, has opened up the swiftly expanding field of Dirac electron optics based on ultraclean ballistic graphene devices.
Correspondingly, optics analogues  comprise Klein 
tunneling in single-layer 
graphene p-n-p junctions \cite{KNG06,Shytov2008,Young2009,Masir2010,Nam2011,Wang2015,Handschin2016}, p-n junctions \cite{CF06,RML+13,Grushina2013}, or Fabry-P\'erot type settings \cite{Handschin2016,KraftPRL2020,Rehmann2019}
as well as anti-Klein tunneling in bilayer graphene \citep{KNG06,VLK+14,Varlet2015,Park2011,RLM+18,Rickhaus2020} where in particular circular p-n junktions were considered \citep{Peterfalvi2009}.
Collimation \cite{CF06,Wang19}, various
electron lensing \cite{CFA07,LGR17,Boggild2017,Brun2019} and guiding \cite{PMPV06,Beenakker2009,Zhang2009,Williams2011,Liu2015,RLM+15,Cheng2019} phenomena were investigated in this context. 

\begin{figure}
\centering
 \includegraphics[width=0.3\textwidth]{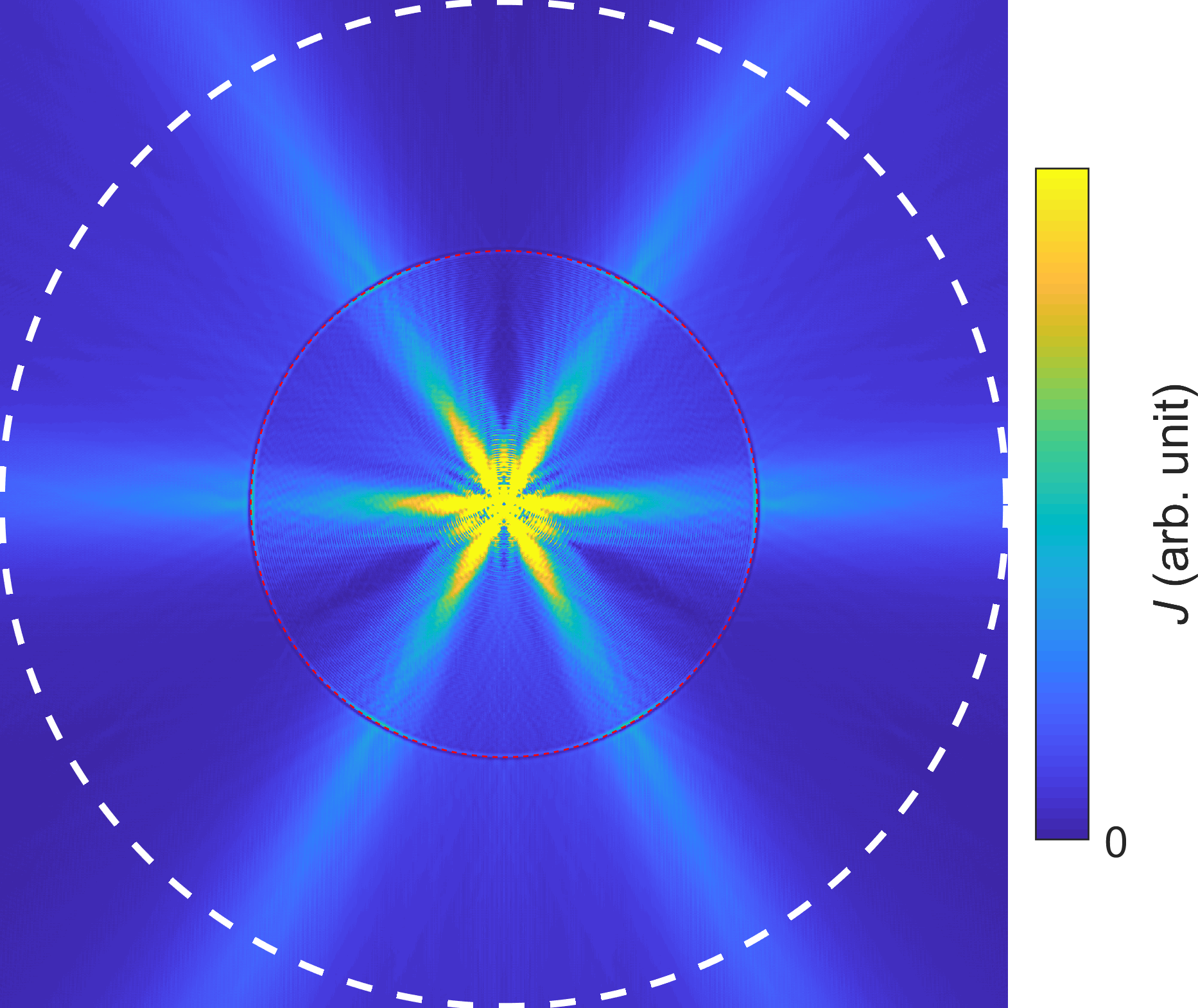}
 \caption{Quantum mechanically calculated local current density for a current injected from a point-like source located at the center of a disk-shaped bilayer graphene cavity (diameter 1$\mu$m), which is attached to four wide transparent leads in order to suppress boundary scattering. 
 The dashed line marks the mid field region ($r_m=2 \mu$m). The hexagonal emission profile reflects the underlying band structure 
 symmetry, see Sec.~\ref{sec_BLGbandstruct}. 
  }
 \label{fig:example}
\end{figure}

\begin{figure}
\centering
  \subfigure[]{\includegraphics[width=0.26\textwidth]{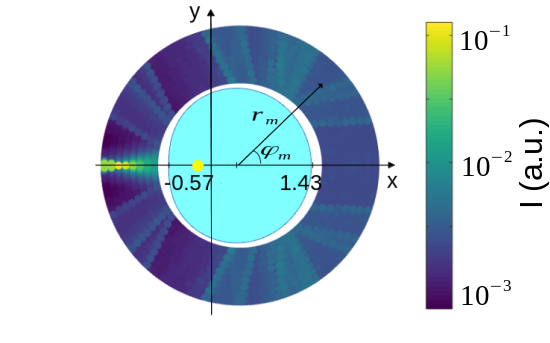}}
  \subfigure[]{\includegraphics[width=0.21\textwidth]{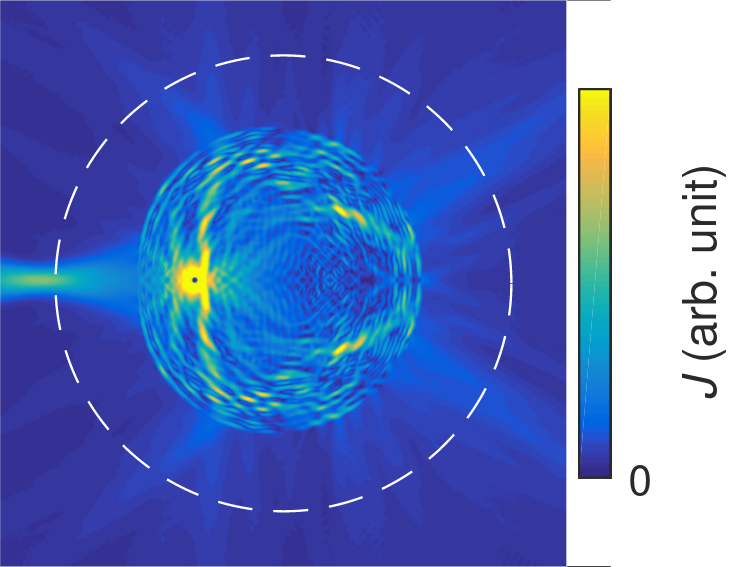}}
  \subfigure[]{\includegraphics[width=0.26\textwidth]{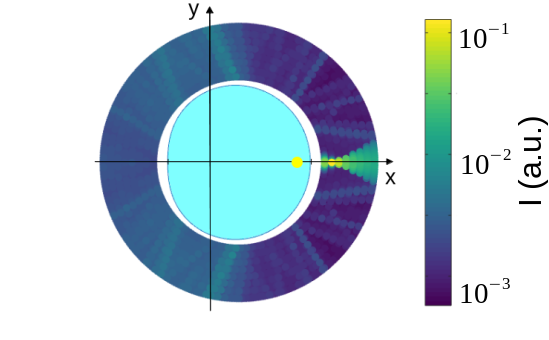}}
  \subfigure[]{\includegraphics[width=0.21\textwidth]{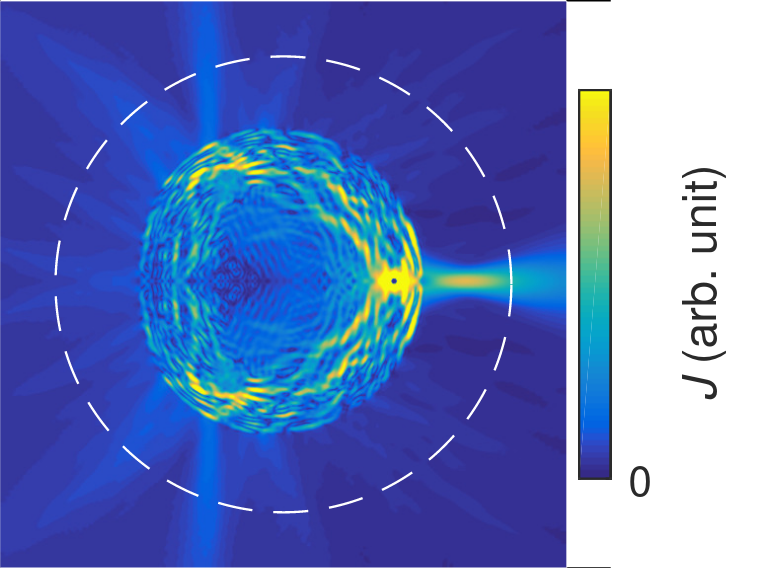}}
 \caption{Particle-wave correspondence in (single-layer) graphene billiards of lima\c{c}on shape in the near and mid field region. 
 The origin of polar coordinates $(r, \phi)$ and mid field coordinates $(r_m, \phi_m)$ is indicated in (a), and a typical mid field boundary $r_m$ (dashed line) in (b,d). 
 (a,c) Particle trajectory simulations, source positions (yellow dot) at $(x,y)=(-0.17,0)$ and $(x,y)=(1.23,0)$, respectively. The light blue area marks the billiards. (b,d) Wave simulations for the same source positions as in (a,c). The color scale in (a,c) shows the total intensity $I$ of summed Fresnel-weighted rays in cells with center 
 $(\phi_m,r_m)$. 
 In (b,d) the color scale is the local electronic current density 
 $J$. 
Classical dynamics and wave results agree semi-quantitatively and both show a pronounced collimation of electrons leaving the cavity to the left (a,b) or right (c,d).
The reason is a lensing effect in this SLG billiards with $n$=-1. }
 \label{fig:lens}
\end{figure}

Complementary to the use of top gates in several of the aforementioned electron steering experiments, recently a scanning tunneling setting has been employed to create disk-like cavities in graphene defined by circular p-n junctions and to probe whispering-gallery type resonant states that are most stable against decay from the cavity via Klein tunneling \cite{ZWN+15}; in a first subsequent theory work non-reciprocity of these whispering gallery modes was predicted \cite{RL15}.  

In earlier theoretical works on differently-shaped, open graphene cavities, the influence of the character -integrable versus chaotic - of the classical charge carrier density dynamics were studied on resonant states in transport \cite{BTB09} and in closed graphene cavities on their sprectra \citep{WAR11}.

The intriguing, novel setup of Ref. \cite{ZWN+15} uses a scanning-tunneling probe to define p-n junction-based billiards. This will allow for experimentally controlling the size and local carrier density of such well-defined ballistic graphene cavities and has been motivating us to consider generalizations of such systems beyond the disk geometry, aiming at charge trapping and controlled directed carrier emission from deformed 
cavities. To this end we adopt and generalize ideas and techniques from the field of optics in photonic micro-cavities. There, corresponding settings for electromagnetic radiation had been successfully used to achieve and control highly directional emission from asymmetrically shaped, lasing cavities \cite{NS97,annbill, reviewletters2008}. Such deformed dielectric microcavity billiards are characterized by an optical refractive index $n$. In these systems light is at least partially confined by total internal reflection in so-called whispering-gallery type modes \cite{annbill}. Breaking of the rotational symmetry was found\cite{reviewletters2008,applphyslett2015} to lead to directional emission from the microcavity. Analyzing the ray-wave correspondence yielded a profound understanding of the behaviour in the optical case based on the nonlinear ray dynamics:
The cavity geometry determines the phase space structure of the rays inside the cavity in the classical, ray limit of optics. This phase space is typically mixed, i.e. comprising co-existing regular and chaotic phase space regions that affect wave-optical emission from the cavity via Fresnel's law. Tuning the ray phase space by deformation of the cavity allowed one for steering directional emission and lasing in the optics context.

Hence it is tempting to explore such a behavior in graphene billiards using a related trajectory-wave correspondence-based approach for electrons in graphene.
Recently, such concepts from mesoscopic optics have been employed for certain single-layer graphene cavity setups: Based on the photonic annular geometry used in Ref.~\cite{annbill} and assuming a ferromagnetic exchange field it was shown in Ref.~\cite{Lai18} that the corresponding internal dynamics of spin-up and -down electrons can strongly differ, leading to specific quantum scattering and polarization features. In Ref.~\cite{Lai18a} the decay features of integrable disk- and chaotic stadium-type cavities were studied based on classical ray tracing.

In the present work we analyze charge carrier trapping and (directed) emission of deformed graphene micro-disks by considering the full particle-wave correspondence through classical and quantum simulations for leaky graphene-based billiards. They are defined by the geometry of a p-n interface that in turn is determined by a gate voltage step from $V_L$ to $V_R$ where $V_L$ is related to $V_R$ by an effective index $n$ of refraction, $V_L=n V_R$. The introduction of a refractive index $n$ is motivated by its aforementioned optical counterpart for deformed dielectric microcavities. The possibility to easily realize negative refractive indices in graphene adds to the fascination of such a study.

Moreover, we also compare single-layer graphene (SLG) and bilayer graphene (BLG) based ballistic cavities exhibiting distinctly different Klein tunneling behavior at their boundaries: Since the p-n-based boundary in SLG exhibits Klein tunneling, quantum states in a disc with predominantly radial excitation are expected to be short lived, while circular "whispering gallery"-type modes should be longer-lived, as for the photonic  analogue. On the contrary, anti-Klein tunneling for certain parameters of the p-n BLG interface implies the opposite transmission characteristics, implying trapped radial "bouncing ball" modes due to suppressed Klein tunneling, a mechanism that does not exist in graphene's optical counterpart.


Photonic and Dirac-fermionic settings also differ in the way resonant states can be created. Optical pumping in the former could be replaced by local charge carrier injection, for example by vertical injection through a point contact on top of the sample~\cite{HFM+15}, in the latter. This amounts to generalize exisiting ray tracing and wave simulations to the case of local (point) sources in the SLG and BLG cavities, where the source position turns out to be particularly  relevant for the emission characteristics. Figure \ref{fig:example} illustrates the peculiar emission profile of such a point source in the center of a BLG disk.

For the specific simulations for a deformed disk we use graphene billiards of the so-called lima\c{c}on shape, see the sketch in Fig.~\ref{fig:lens}(a).
For the optical case, a robust and resonance-independent directional far-field emission was observed for both wave and ray simulations \cite{reviewletters2008}. 
The corresponding investigation and explanation in the case of graphene is one main subject of this paper that is organized as follows. In Sec.~\ref{sec:raymod} we introduce the trajectory modelling of graphene billiards by introducing Fresnel's and Snell's law for graphene p-n interfaces in Sec. \ref{subsec:Fresnel}  and apply it to ray-tracing simulations in Sec. \ref{subsec:raytracing}, while we comment on the wave simulations in Sec. \ref{subsec_wavesim}. In Sec.  \ref{sec:raywavesingle} we compare our ray and wave simulations results and discuss ray-wave correspondence for SLG billiards. We extend this concept to BLG cavities and highlight the ray-dynamic origin of the different behaviour in Sec. \ref {sec:bilayer} and summarize our findings in Sec. \ref{sec:concl_summary}.

\begin{figure}
  \subfigure[]{\includegraphics[width=0.23\textwidth]{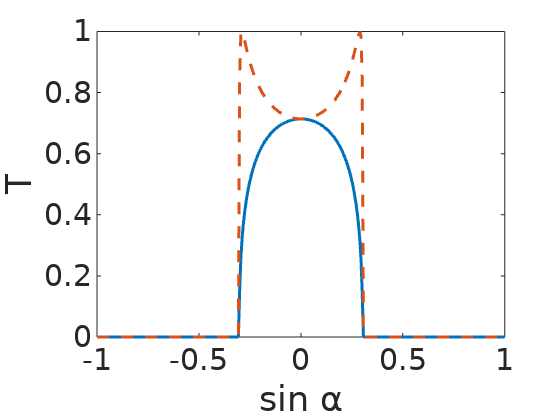}}
  \subfigure[]{\includegraphics[width=0.23\textwidth]{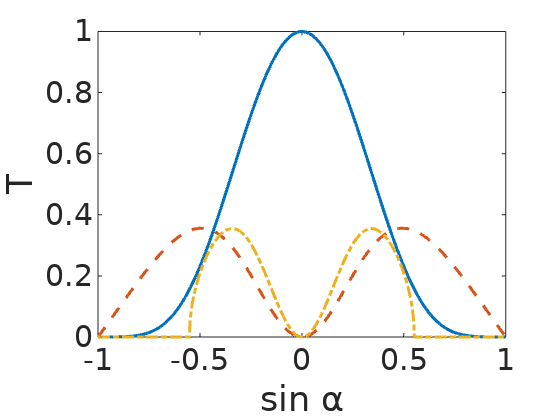}}
 \caption{Fresnel transmission coefficient $T = 1-R$ , with $R$ the reflection coefficient, for (a) the optical case, $n=3.3$ and TM (full line) and TE (dashed line, Brewster angle feature at $T=1$) polarisation. (b) Single-layer graphene, $n=-1$ (full line, smoothed potential), bilayer graphene $n=-1$ (dashed line) and $n=-3$ (light dotted line). A steplike potential and a voltage $V_R $ =20 meV was used for the bilayer graphene data.} 
 \label{fig:fresnel}
\end{figure}


\section{Electron-trajectory modelling of graphene billiards}
\label{sec:raymod}

\begin{figure}
 \subfigure[]{\includegraphics[width=0.45\textwidth]{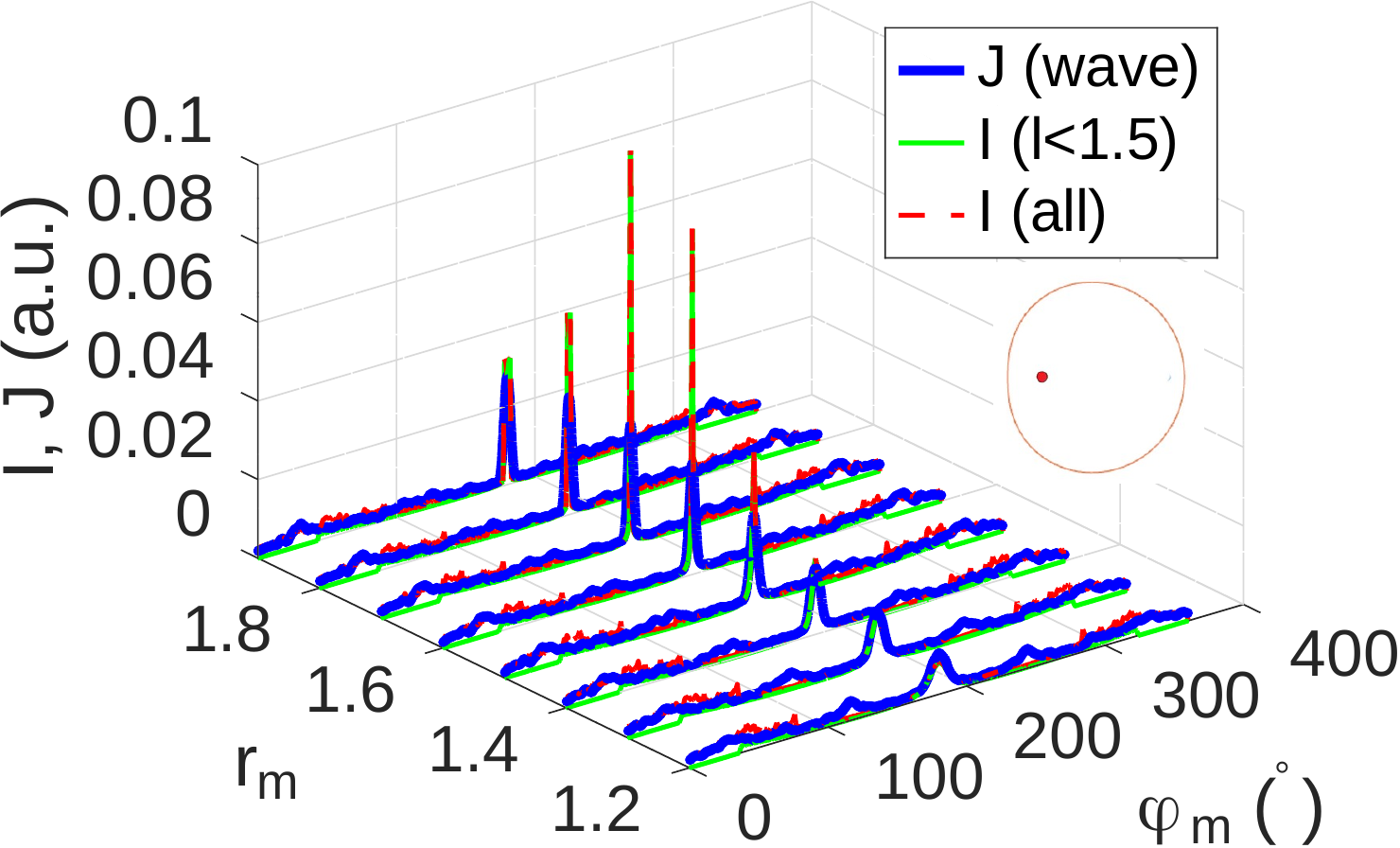}}
 \subfigure[]{\includegraphics[width=0.45\textwidth]{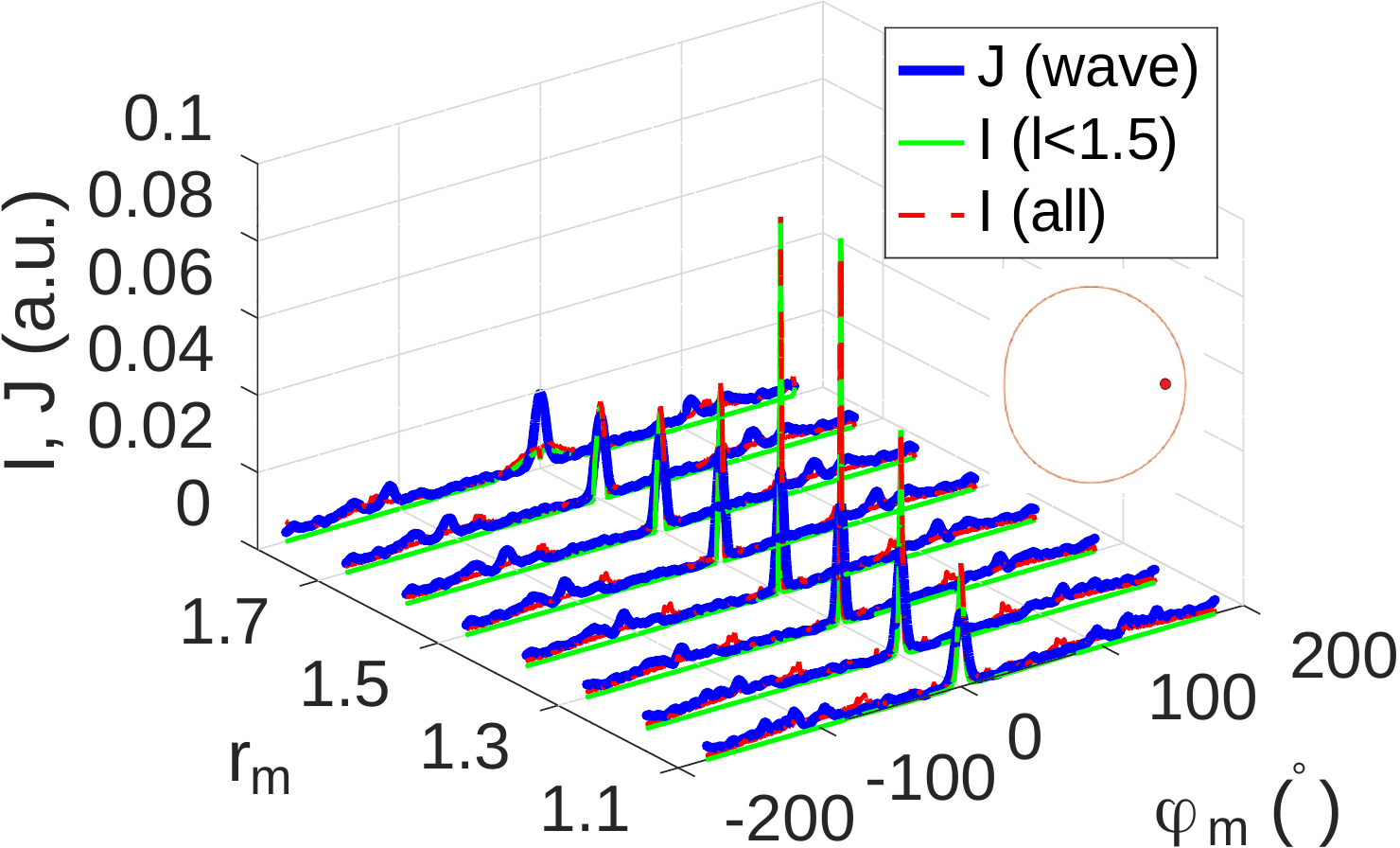}}
 \caption{Lensing effect in the mid field ray intensity $I (r_m, \phi_m)$ 
 for SLG billiards with effective index of refraction $n=-1$.
 The contributions of 
 very short rays ($l<1.5$, green/light full lines) and all rays (red dashed lines) are distinguished and shown with wave simulation results $J$ (blue/dark full lines).  (a) Source position at $x_s=-0.17$ (see inset). The position of the highest peak ($r_m=1.3 \pm 0.1 $) in ray and wave simulations is almost equal to the analytically calculated position $r_f=1.24$. (b) Same for 
 $x_s=1.23$. 
 Again, 
 $r_m=1.7 \pm 0.1 $ and $r_f=1.61$ correspond well.}
 \label{fig:lensfull}
\end{figure}

\subsection{Fresnel's and Snell's law for graphene}
\label{subsec:Fresnel}

We model the graphene cavities as so-called lima\c{c}on billiards that allow for tuning the deformation based on disk-shape and have proven convenient in the photonic case~\cite{limacon_NJPhys}. 
In polar coordinates $(r, \phi)$ the billiard boundary reads $r(\phi) = R_0 (1 + \epsilon \cos \phi$) with mean radius $R_0$ and the deformation parameter $\epsilon$ set here to 0.43. For the optical case, a robust and resonance-independent directional far-field emission was observed for both wave and ray simulations, as well as in experiments \cite{reviewletters2008, limacon_Cao2009, limacon_Kim2009, limacon_Susumu_Taka2009, limacon_NJPhys}. 

A close correspondence between optical billiards for light and graphene billiards for electrons can be established by generalizing Fresnel's and Snell's law to graphene interfaces, cf.~Fig.~\ref{fig:fresnel}. While in the optical case a step in the refractive indices defines a dielectric cavity (typically with refractive index $n>1$ embedded in air with $n_0=1$), a p-n step defines the graphene billiard interface. Here, the dominating feature is Klein tunneling \cite{KNG06,CF06,Klein_ZPhys}.
It yields perfect transmission $T=1$ for electrons with normal incidence onto the interface in SLG, in contrast to $T = T(n) =  1 - ((n-n_0)/(n+n_0))^2 $ in the optical counterpart. 
BLG adds even more variability to Fresnel's law 
with the realization of anti-Klein tunneling ($T=0$ at normal incidence) \cite{KNG06,VLK+14} and, moreover, the possibility to realize the transition from anti-Klein to Klein tunneling \cite{RLM+18}.

Here, we use Fresnel laws for single- and bilayer graphene  
that are obtained from numerical calculations of the angle-resolved
transmission function across a smooth p-n junction  
based on the real-space Green's function approach\cite{Datta1995} within the tight-binding framework. See Ref.\ \onlinecite{Liu2012} for technical details.
The resulting Fresnel laws are shown in Fig. \ref{fig:fresnel} and were numerically implemented as boundary conditions for our classical ray simulations used below where wave propagation is approximated by ray tracing of electron trajectories inside the billiards. 
We point out that the effect of the electron's spin can be neglected in the present study as the spin-orbit coupling effect in graphene is of the order of only several micro eV such that the Fresnel laws will practically be the same for both spin species.

Similar to Fresnel's law, Snell's law can be generalized to the graphene case as well. It reads 
\begin{equation}
 \frac{V_L}{V_R}
\sin \alpha= n\sin \alpha = \sin \beta \:,
\label{eq:snellgraph}
\end{equation}
where $\alpha$ is the angle of incidence of a light ray inside the cavity and $\beta$ is the angle of the refracted ray leaving the cavity. Both are measured with respect to the boundary normal. The voltages $V_L$ and $V_R$ define the potential step and are related to the Fermi energy $E_F$ and the potential step height $V_0$ by $E_F = - V_L$ and $V_0 = V_R - V_L$, yielding
\begin{equation}
 n=\frac{E_F}{E_F - V_0}
\label{eq:n}
\end{equation}
 as an alternative expression for the effective refractive index $n$. 
In the following, 
values $n=-1$ and $n=-3$  are  used ($n_0=1$). Note that the Fresnel transmission coefficient may depend not only on the voltage ratio but also on the voltage value itself as in the case of BLG. 

SLG and optical billiards show, qualitatively, a similar behaviour of the transmission coefficient $T$ as a function of the angle of incidence $\alpha$, cf.~Fig.~\ref{fig:fresnel}. In particular, and despite the presence of Klein tunneling, the transmission $T$ drops to zero as $\alpha$ reaches grazing incidence, just as known for total internal reflection at optical interfaces for angles larger than the critical angle $\alpha_c = \arcsin n_0/n$. Consequently, we expect a certain analogy in (the interpretation of) the behaviour of optical and SLG billiards, concerning e.g. the importance of whispering-gallery type resonances.

BLG shows, in general, smaller transmission coefficients $T$ with zero transmission $T=0$ at normal incidence, cf.~Fig.~\ref{fig:fresnel}(b). Therefore, the confinement of electrons in BLG billiards is expected to be better than in the SLG case, in particular for higher effective refractive indices. This will be important when interpreting ray modelling results of BLG vs. SLG electron sources below.

\subsection{Trajectory simulations}
\label{subsec:raytracing}

Particle trajectory simulations according to SLG and BLG Fresnel's and Snell's law (Fig.~\ref{fig:fresnel} and Eq.~(\ref{eq:snellgraph}))
have been performed numerically. 
To mimic the electronic source, 4000 ray-like trajectories were started at the position of the point-like source in random directions. Their trajectories and intensity evolution across the graphene billiards
were followed, and each of them was reflected at the p-n interface up to 70 times or until its intensity $I$ dropped below $5 \times 10^{-6}$ of the initial intensity $I_0$. Every time a particle is reflected at the p-n step, Snell's and Fresnel's law for graphene interfaces 
are used to calculate the angle of refraction and the reflected and refracted intensities. 
The current density in mid field is calculated using the number of trajectories crossing one of the 360 cells in mid field at radius $r_m$, weighted by their Fresnel intensity. Note that the origin of mid field polar coordinates $(r_m, \phi_m)$ is in the apparent center of the billiards, cf.~Fig.~\ref{fig:lens}(a), and differs from the origin of the $(r, \phi)$ coordinates used to describe the billiards boundary.

Ray simulations for optical billiards in the so-called stationary regime have proven to be a useful and reliable concept to describe the far field emission characteristics of lasing microcavities \cite{applphyslett2015, limacon_NJPhys}, and generalization to mid fields is straightforward. 
Note, however, that here a source-driven billiards has to be considered in contrast to uniformly pumped microcavity lasers in the stationary regime. In the ray picture, the latter are described by long trajectories that have reached the stationary regime, which is 
an exponential decrease of the total intra-cavity intensity. Then, the far field emission characteristics is determined by the so-called natural measure (or Fresnel-weighted unstable manifold or steady probability distribution) in phase space~\cite{Lee2005,limacon_NJPhys, reviewletters2008, applphyslett2015}. We point out that the transient regime occurring prior to reaching the stationary regime 
had to be abandoned in these cases. It is characterized by faster-than-exponential decay due to loss of all (randomly and with unit intensity started) trajectories that are not confined by total internal reflection and leave the billiards  ``right away''. Once the trajectories have adopted the system-specific intensity distribution, that is, have arrived at the natural measure, the correct far field characteristics could be obtained. 


In the presence of sources, new particles (or rays) are constantly being fed into the system and, therefore, short paths cannot be neglected. So simulations for billiards with sources have to be extended by the short trajectories that were left out before. To this end, we divide the paths contributions in those from  short trajectories with $l<1.5$ or $l<3$, intermediate long trajectories with $3<l<20$, and long trajectories $l>20$ when discussing the results of ray simulations in the following sections. All lengths are measured in units of the mean cavity radius $R_0$ (that is typically set $R_0=1$). Source positions within the cavity are described in $(x_s,y_s)$ coordinates (same origin as $(r, \phi)$) while sources along the cavity boundary are given through their polar angle $\phi_s$ (same origin as $(r_m, \phi_m)$).

\begin{figure}
\includegraphics[width=\columnwidth]{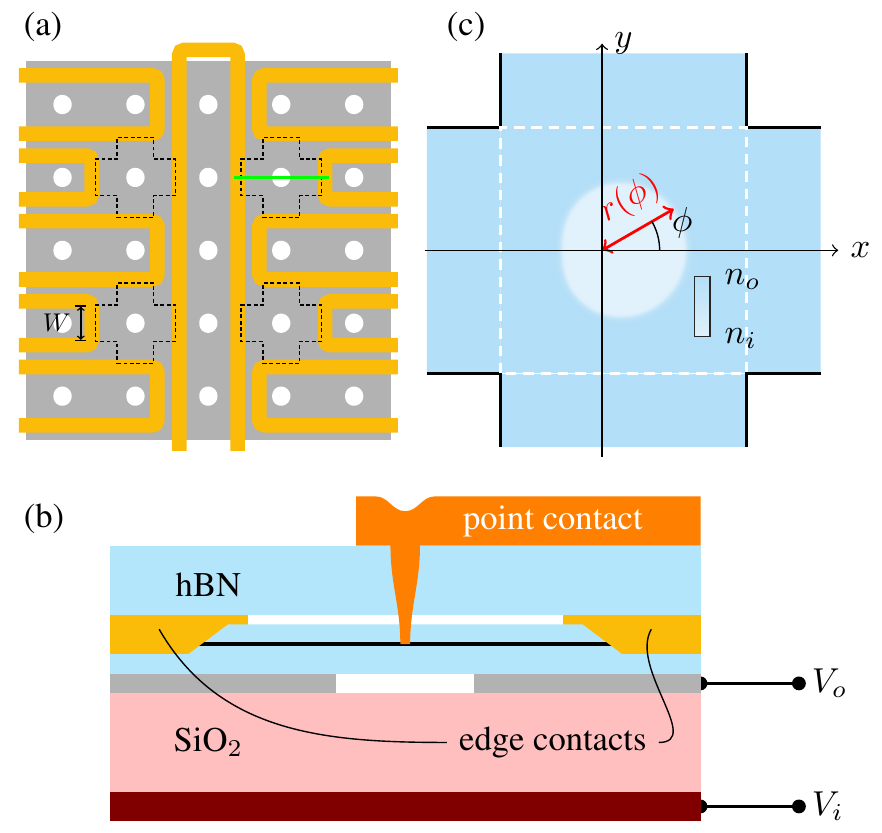}
\caption{ Numerical wave simulation of graphene billiards. (a) Top view of periodically etched bottom gates (gray) and the edge contacts (yellow). The dashed cross bars mark four graphene samples attached to four contacts each, one of the cross bars is magnified in (c). The white dashed square in (c) indicates the considered scattering region for quantum transport simulations and is attached to four semi-infinite leads of width $W$. 
(b) Possible experimental realization of a graphene device containing the lima\c{c}on cavity, showing the cross-sectional side view along the green line in (a) and indicating the point contact (orange), hexagonal boron niride (hBN) dielectric layers (light blue), graphene sample (black), the holey bottom gate (gray, $V_o$), the SiO$_2$ substrate (pink), and the global back gate (brown, $V_i$).}
\label{fig:device}
\end{figure}

\subsection{Wave simulations}
\label{subsec_wavesim}

For wave simulations we consider a square-shaped scattering region attached to four semi-infinite leads of width $W$ to model the cross bar depicted in Figs.~\ref{fig:device}(a,c). The cross bar can be made of SLG and BLG, both of which can be straightforwardly described by tight-binding models in real-space. To consider the experimentally feasible size ($W=2~\mu\text{m}$ and mean lima\c{c}on radius $R_0=0.5~\mu\text{m}$ in all wave simulations), we adopt the scalable tight-binding model\cite{Liu2015} with the scaling factor $s_f=8$ for SLG and $s_f=4$ for BLG. In addition to the four planar semi-infinite leads that work as drain terminals, the point-like injector is modeled by a vertical lead with a disk cross section of radius about 24~nm, adopting the same method used in Ref.\ \onlinecite{LGR17}.

We compute the spatially resolved quantum-mechanical probability current density using two different but consistent methods. For SLG, we apply the nonequilibrium Green's function method\cite{Datta1995} in the limit of equilibrium using our own code, same as Ref.\ \onlinecite{LGR17}. For BLG, we use the open-source code \textsc{KWANT} \cite{Groth2014} to compute the local current density. All wave simulations are done at zero temperature.

The setup modelled is closely related to the experimental realization of a lima\c{c}on shaped graphene billiards, cf.~Fig.~\ref{fig:device}(b). Due to screening, the global backgate applied with gate voltage $V_i$ influences only the central part of the graphene sample inside the lima\c{c}on cavity, while the holey bottom gate applied with gate voltage $V_o$ controls the region outside the cavity. Combination of gate voltages $V_i$ and $V_o$ allows us to electrostatically and controllably induce at the center of the graphene sample a 
lima\c{c}on-shaped cavity. At the boundary of the cavity, the gate-induced carrier density is expected to smoothly transition from $n_i$ in the inner region to $n_o$ in the outer region, with the smoothness depending mainly on the thickness of the bottom hexagonal boron nitride (hBN) layer. The carrier density profile is exemplarily shown based on a model function in Fig.\ \ref{fig:device}(c).
All wave simulations are based on such a four-terminal structure, considering $W=2~\mu\text{m}$, $R_0=0.5~\mu\text{m}$, and $n_i=-6\times 10^{11}~\text{cm}^{-2}=-n_o$. In addition, we will assume reflectionless contacts such that we may model the cross bar by considering a square scattering region marked by the white dashed box in Fig.\ \ref{fig:device}(c)] attached to four semi-infinite leads. In ray simulations, on the other hand, the system is further simplified to a boundless graphene sample, to be explained in \autoref{subsec:raytracing}.

\section{Particle-wave correspondence in single-layer graphene billiards}
\label{sec:raywavesingle}

\begin{figure}
\centering
  \subfigure[]{\includegraphics[width=0.42\textwidth]{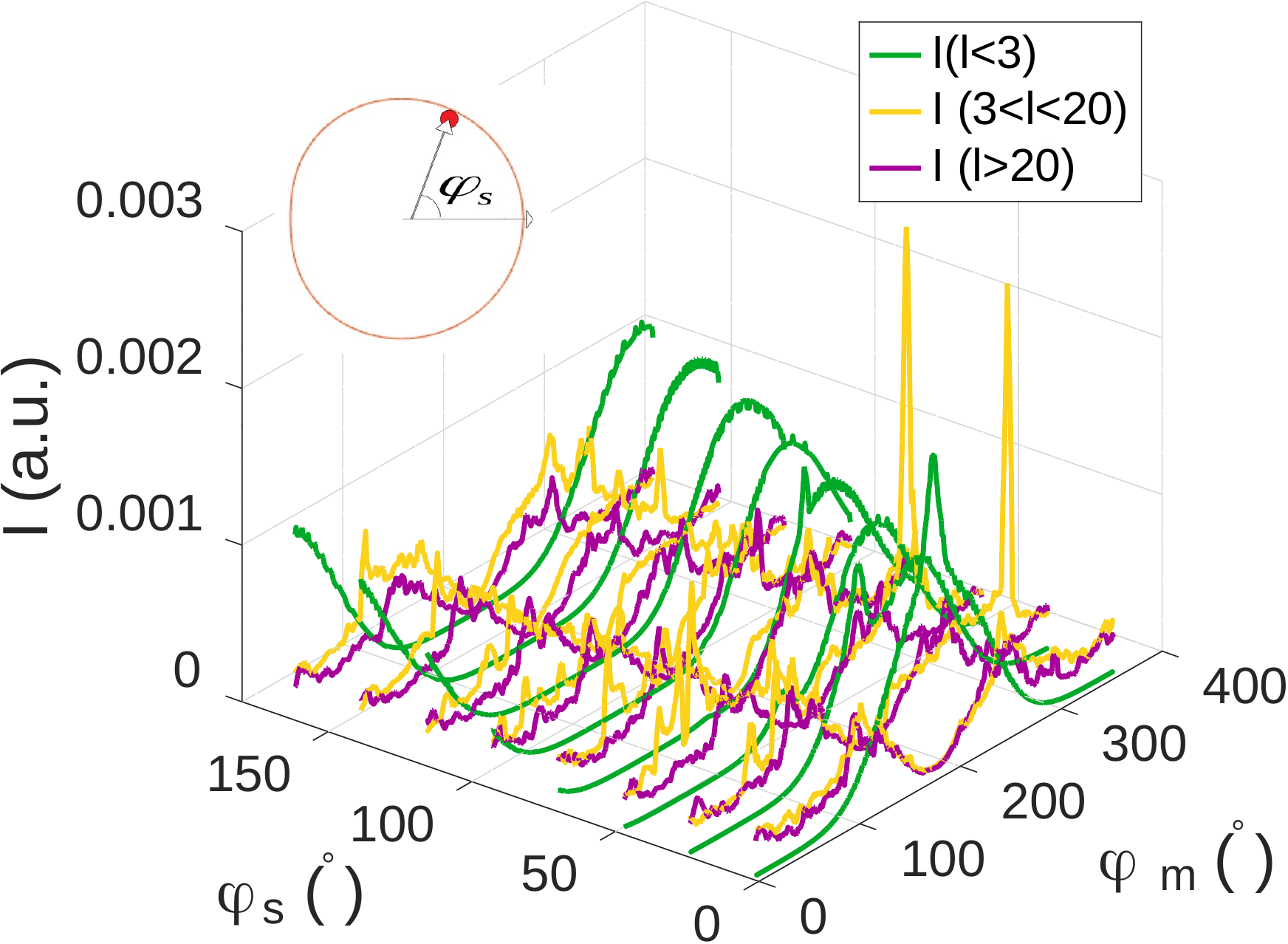}}
  \subfigure[]{\includegraphics[width=0.42\textwidth]{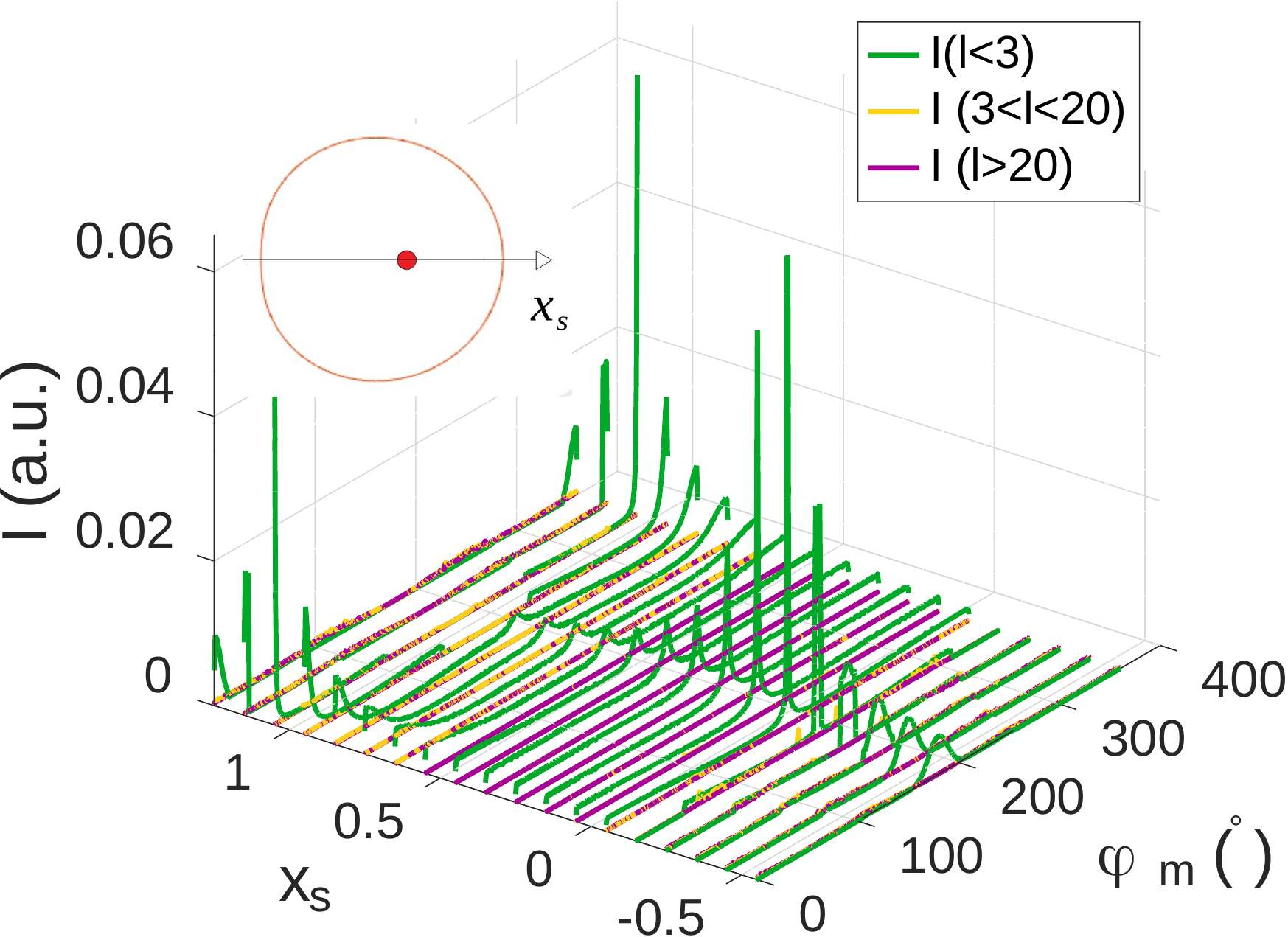}}
  \subfigure[]{\includegraphics[width=0.42\textwidth]{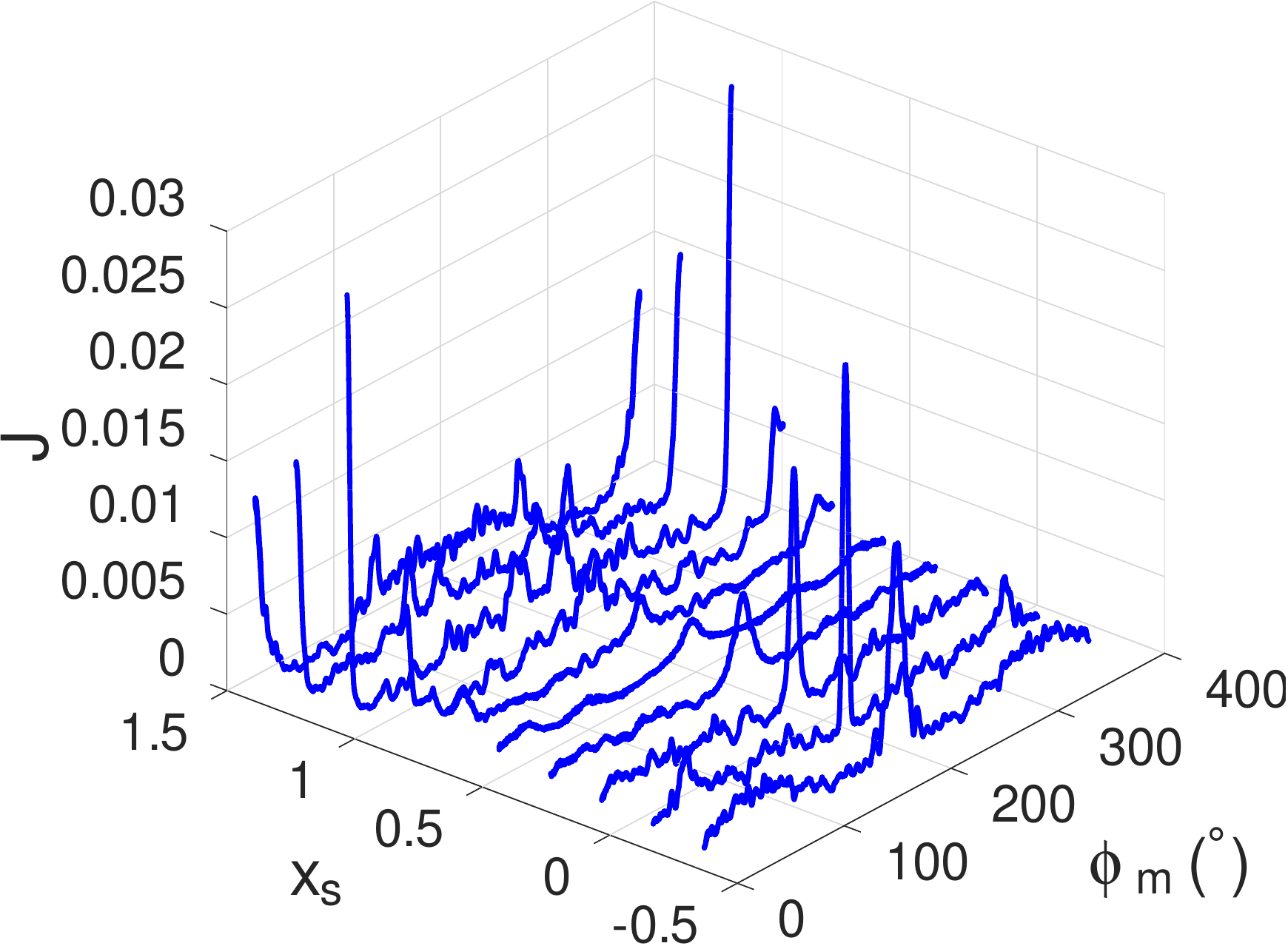}}
 \caption{SLG billiards trajectory intensity $I (r_m=2,\phi_m)$ for varying source position $x_s$ 
 and paths of different lengths: short paths $l<3$ (green/grey full line), intermediate length paths $3<l<20$ (yellow/light grey) and long paths $l>20$ (purple/dark grey) for
 (a) different source positions $\phi_s$ along the billiards boundary
 and (b) different source positions $x_s$ along the x-axis (with $y_s=0$). 
 (c) Same as (b) but wave simulations results for the wave intensity $J$  which shows, qualitatively, a very similar behaviour.} 
 \label{fig:edge}
\end{figure}

\subsection{Lensing effect}

For source positions $x_s$ along the $x$-axis (symmetry axis of the cavity, $y_s=0$), a semi-quantitative agreement between trajectories and wave simulations can be observed as shown in Fig.~\ref{fig:lens}. Both intensity distributions show a very pronounced peak in mid field emission whose position depends on the source position $x_s$. We now investigate this behaviour in detail by scanning the source position $x_s$ along the $x$-axis. 

The results are presented in Fig.~\ref{fig:lensfull}. In order to resolve the peak evolution, the mid field observation radius $r_m$ was varied.
Clearly, a pronounced peak in the emission intensity develops in a tiny range of mid field radii $r_m$. Its dominating emission directionality $\phi_m$ depends on the source position $x_s$; see Fig.~\ref{fig:lensfull}(a,b). 
Note that this peak is mainly caused by short rays with trajectory lengths $l<1.5$. This finding reflects the fact that the SLG system is very open because, according to its Fresnel law, rays with normal incidence onto the p-n step are fully transmitted, and confinement by total internal reflection is only reached for relatively large angles of incidence $\alpha$. 

While the appearance of pronounced preferred emission directions in the wave simulations, Fig.~\ref{fig:lens} (b),(d), might on first sight suggest a coherence effect that cannot be captured by a naive ray model description, it turns out that the opposite is true: the ray model nicely explains it as a lensing effect of the (almost) perfect graphene lens with $n=-1$ as we shall see now. 

Using the approximation for rays near the optical axis \cite{hechtoptics}, we find the following relation for a thick graphene lens of refractive index $n=-1$, 
\begin{equation}
\frac{1}{f}=\frac{1}{s}+\frac{n-1}{r_c}\,
\end{equation}
where $r_c$ is the (local) radius of curvature of the cavity near the source ($r_c=2.3$ for $\varphi=\pi$ and $r_c=1.1$ for $\varphi=0$), $f$ is the distance of the lens' focal point to the edge of the cavity and $s$ is the distance between the source and the edge. The analytical results for the position of the focal point $r_f = f+ R_0 = f+1 $ correspond very well to peak positions obtained in the trajectory and wave simulations; cf.~Fig.~\ref{fig:lensfull}.

\begin{figure}
\centering
 \subfigure[]{\includegraphics[width=0.45\textwidth]{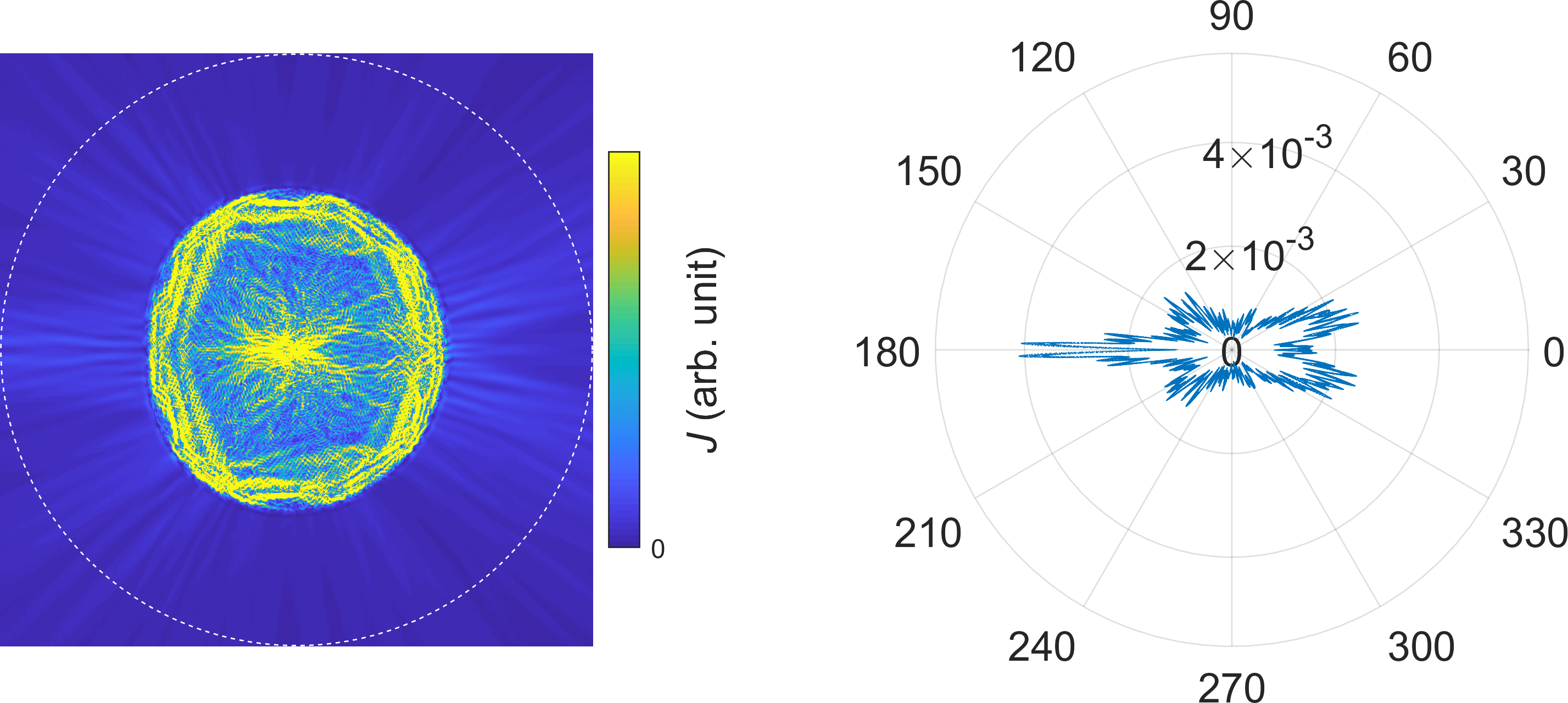}}
  \subfigure[]{\includegraphics[width=0.45\textwidth]{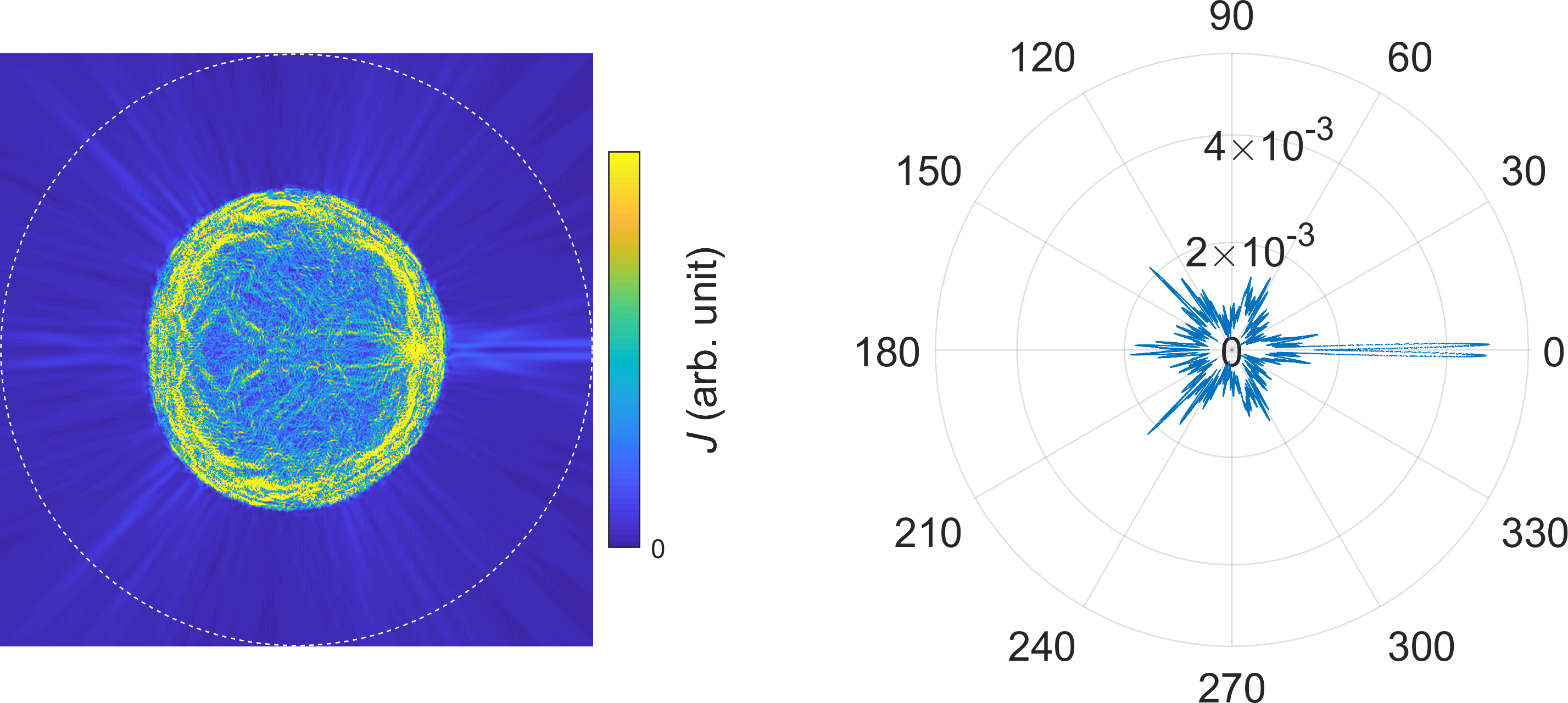}}
  \subfigure[]{\includegraphics[width=0.46\textwidth]{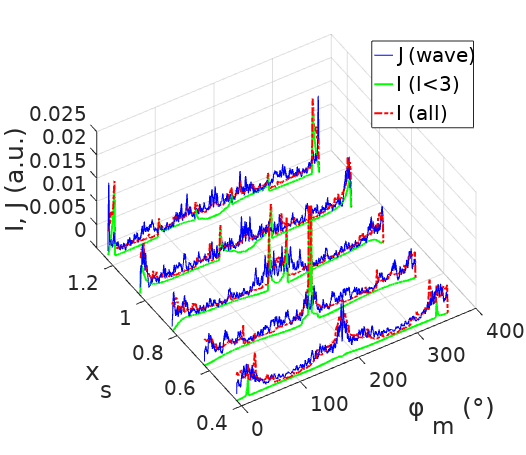}}
 \caption{Particle-wave correspondence in BLG billiards. Wave resonances in BLG for source positions (a) $x_s=0.43$ and (b) $x_s= 1.23 $ (left panels) and corresponding mid field emission as polar plot (right panels). The anti-Klein tunneling prevents trajectories emerging from the source and hitting the boundary at normal incidence to directly leave the cavity, in contrast to SLG. Rather, a whispering-gallery type mode along the system's boundary can form in BLG. 
 (c) Mid field intensity into direction $\phi_m$ for wave simulations (blue/dark solid line, $V_R = 100$ meV, $n$=-1) and the corresponding ray simulation result (short rays (all rays) indicated by the green/grey solid (dashed) line. 
 for source position $(x_s, y_s=0)$ 
 and mid field radius $r_m=2$. We find semi-quantitative agreement and the main emission directions well reproduced within the classical model, in particular by long paths. 
 }
 \label{fig:raywavebilayer}
\end{figure}

\subsection{Varying source positions}

Next, we want to investigate the influence of the source position on the formation of the lensing effect in more detail. More generally, the objective is to characterize the dynamics of the ray model for arbitrary source positions by discussing the contributions from the transient and the stationary regime, or shorter and longer paths, respectively. The results are summarised in Fig.~\ref{fig:edge} for SLG billiards with 
$n=-1$.

In Fig.~\ref{fig:edge}(a), the source position $\phi_s$ is varied along the billiard's boundary.
Note that electrons 
have to be reflected 
at least once before transmission. The main emission originates from rays that have crossed the billiards just once and leave opposite the source. Shifting the source position continuously from $\phi_s=0^o$ to $180^o$ yields a smooth shift of the mid field emission peak. 
We point out that the emission mechanism is very different from the lensing effect discussed before.

This source-related main emission peak is dominated by the short paths 
 (green) 
curve in Fig.~\ref{fig:edge}(a). However, all longer trajectories (purple) 
contribute to a background with characteristic mid field emission directions $\phi_m=120^o$ and $240^o$. These can be explained by the Fresnel-weighted unstable manifold
for graphene billiards (see Fig.~\ref{fig:poincare} below). They correspond to the stationary, source independent emission characteristics of the system. Notice that the long-path results are practically independent from the source position and are symmetric \cite{reviewletters2008,limacon_NJPhys}, even if the source is not placed at the x-axis. 

In Figs.~\ref{fig:edge}(b) and (c), the source position is varied along the $x$-axis. We compare the trajectory simulation result (b) and the wave simulation result (c) and find very close and reasonable agreement. The dominance of short path contributions (green line) to the overall intensity is clearly visible. Around $x_s=0.5$, that is, when the source crosses the apparent center of the lima\c con cavity (i.e., the origin of the $(r_m, \phi_m)$ coordinates), the output peak related to the lensing effect switches from $\phi_m = 180^o$ (for smaller $x_s$) to $\phi_m = 0^o$ as illustrated in Figs.~\ref{fig:lens} and \ref{fig:lensfull}. We point out that for central source positions the emission is almost uniform and does not reflect the broken rotational symmetry of the cavity. 
The reason is that central source positions induce predominantly  normal incidence onto the boundary 
such that Klein-tunneling initiated perfect transmission carries most of the ray intensity outside within very few reflections. Consequently, the peak intensity is smallest at central source positions around $x_s=0.5$, cf.~Figs.~\ref{fig:edge}(b) and (c). For source positions closer to the boundary, larger angles $\alpha$ of incidence become typical and support longer trajectories. 


\section{Bilayer graphene billiards}
\label{sec:bilayer}

\subsection{Particle-wave correspondence}

\begin{figure}
\centering
 \subfigure[]{\includegraphics[width=0.42\textwidth]{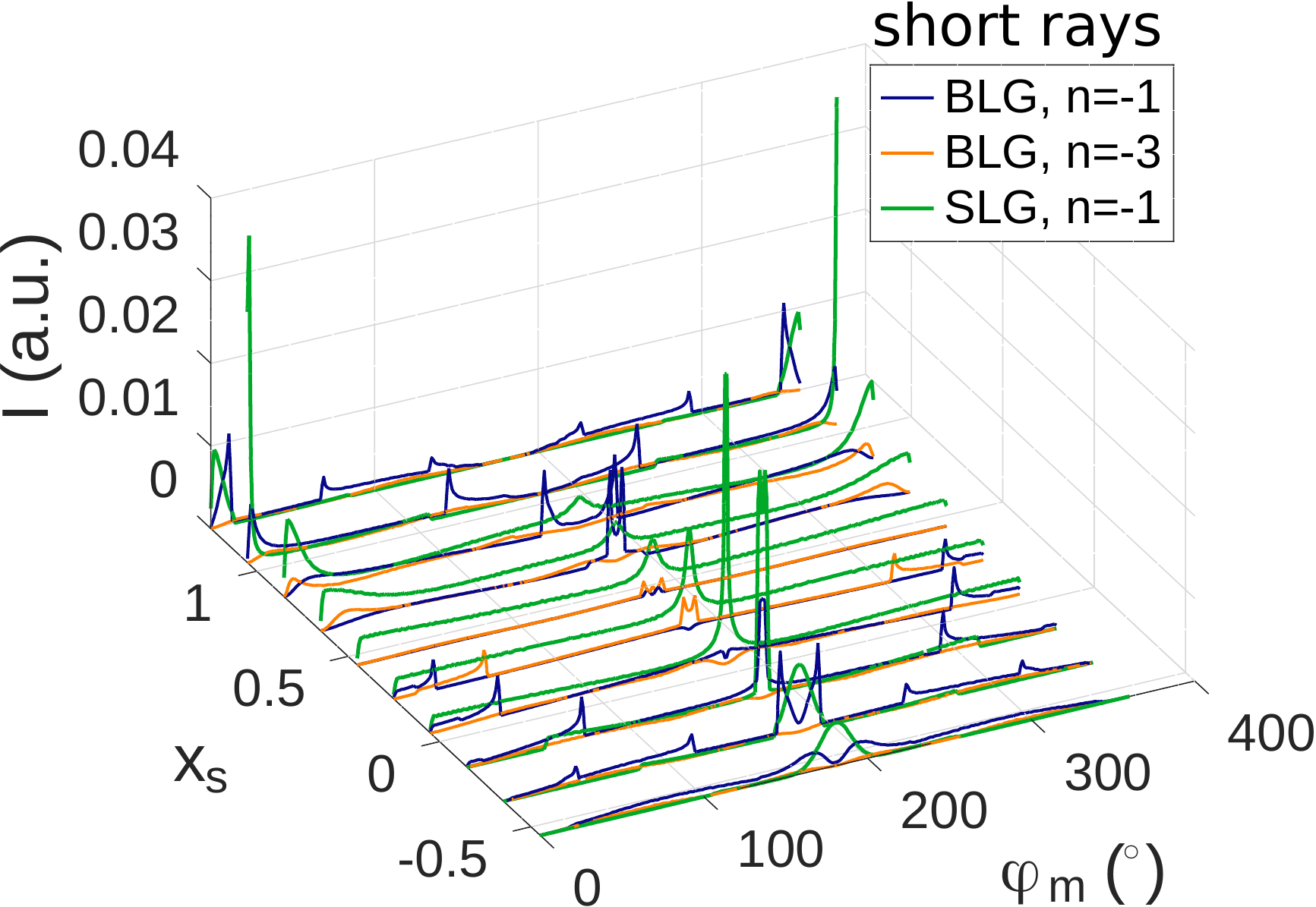}}
 \subfigure[]{\includegraphics[width=0.42\textwidth]{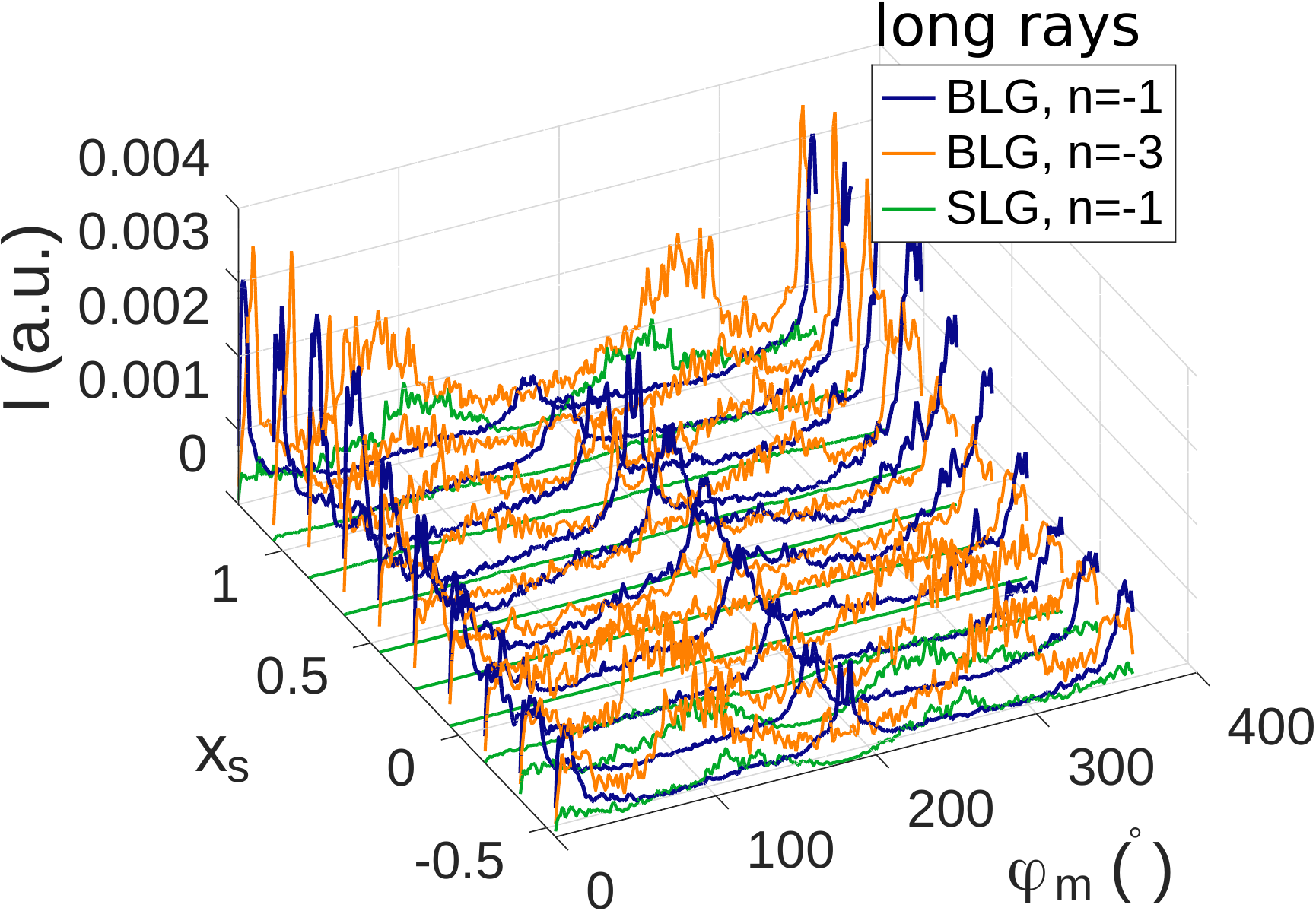}}
 \subfigure[]{\includegraphics[width=0.42\textwidth]{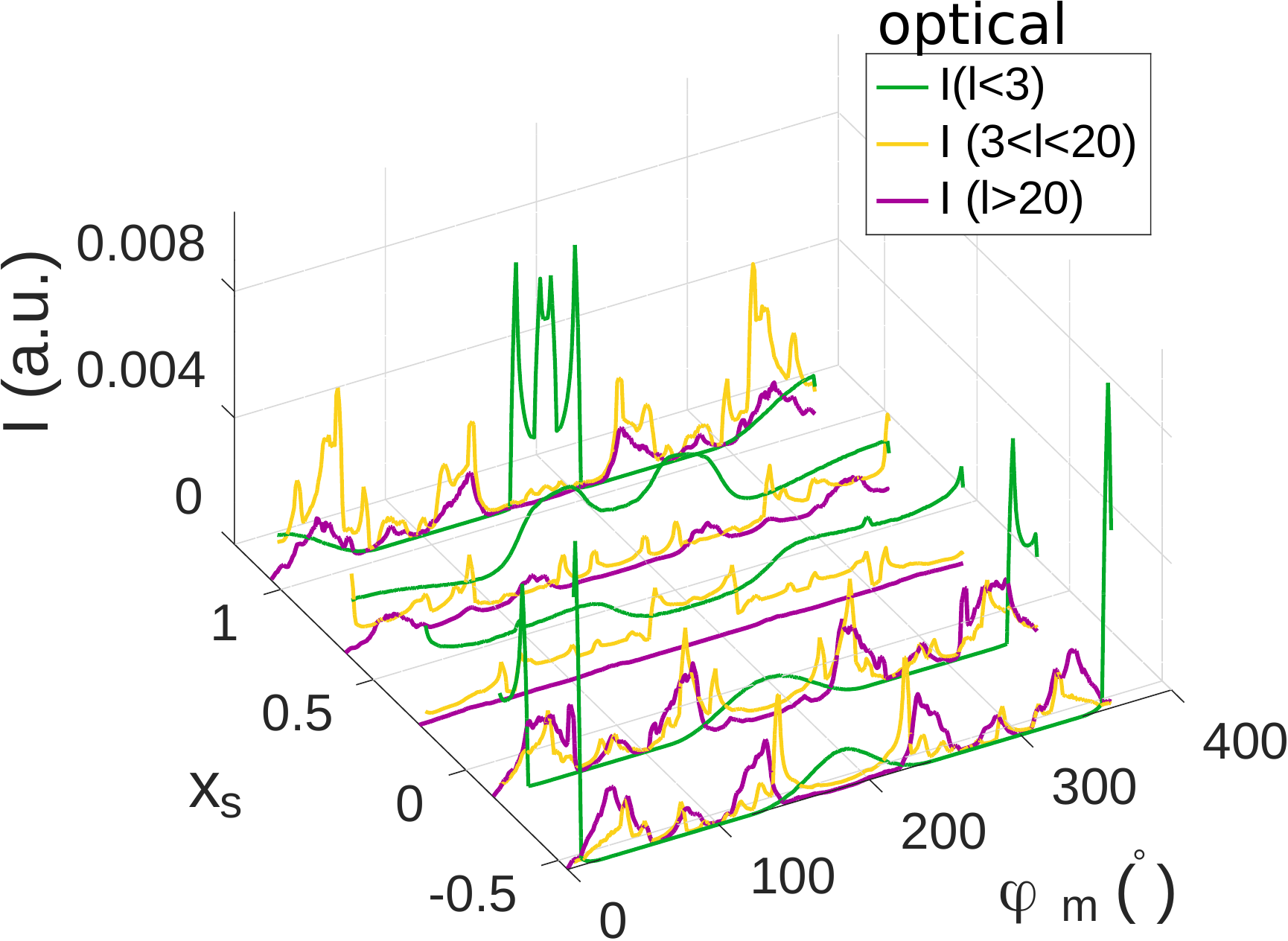}}
 \caption{(a) Short and (b) long trajectory contributions to the mid field emission characteristics $I (r_m=2, \phi_m)$ of bilayer ($V_R=20$ meV, $n=-1$, green and $n=-3$, orange) and single-layer ($n=-1$, purple)  graphene billiards for varying source position $(x_s,y_s=0)$. (c) Optical case, TE polarization, for comparison.
 Short orbits dominate the emission especially for SLG in (a), while long orbits determine the emission characteristics of BLG in (b), reflecting the stronger confinement of electrons in BLG. 
  }
 \label{fig:singlelayerbilayer}
\end{figure}

\begin{figure}
\centering
  \subfigure[]{\includegraphics[width=0.4\textwidth]{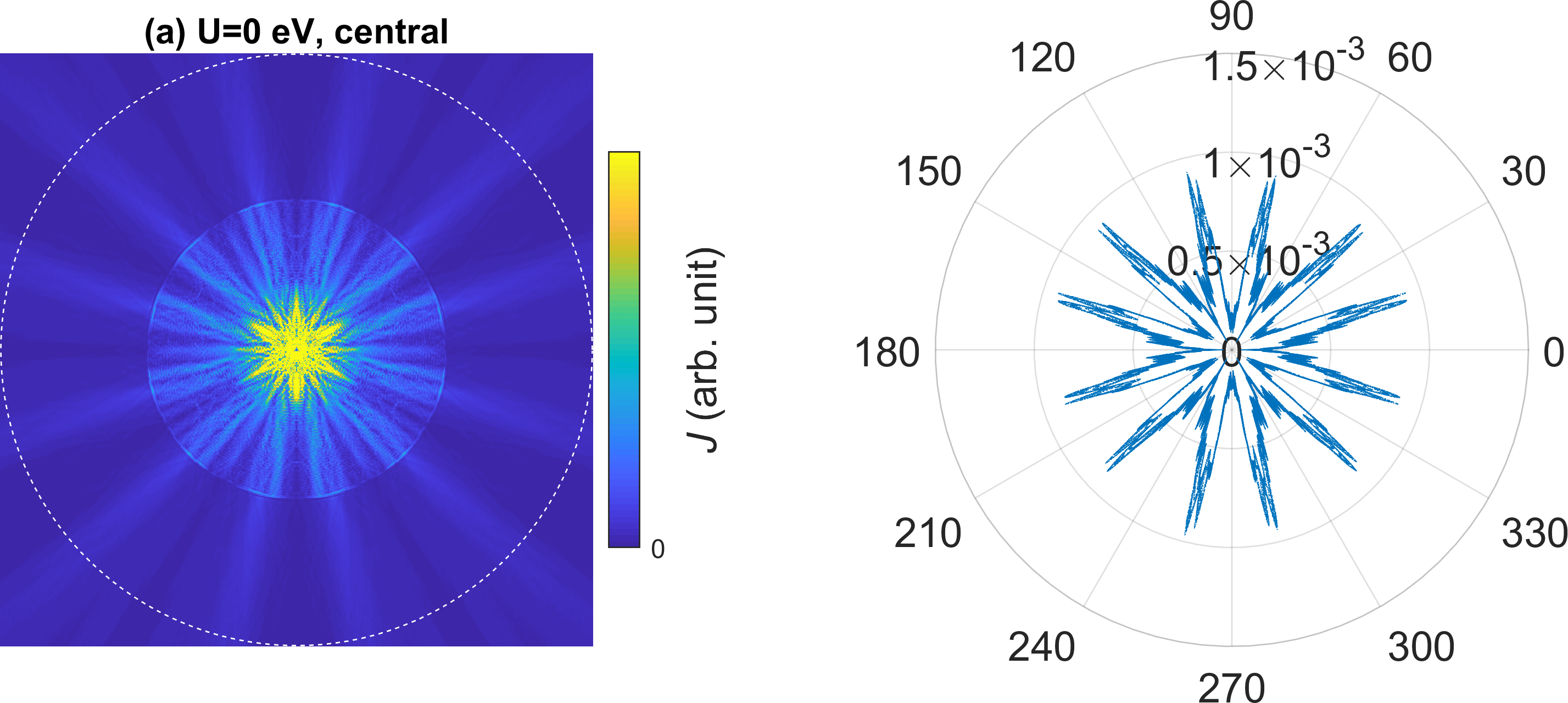}}
  \subfigure[]{\includegraphics[width=0.4\textwidth]{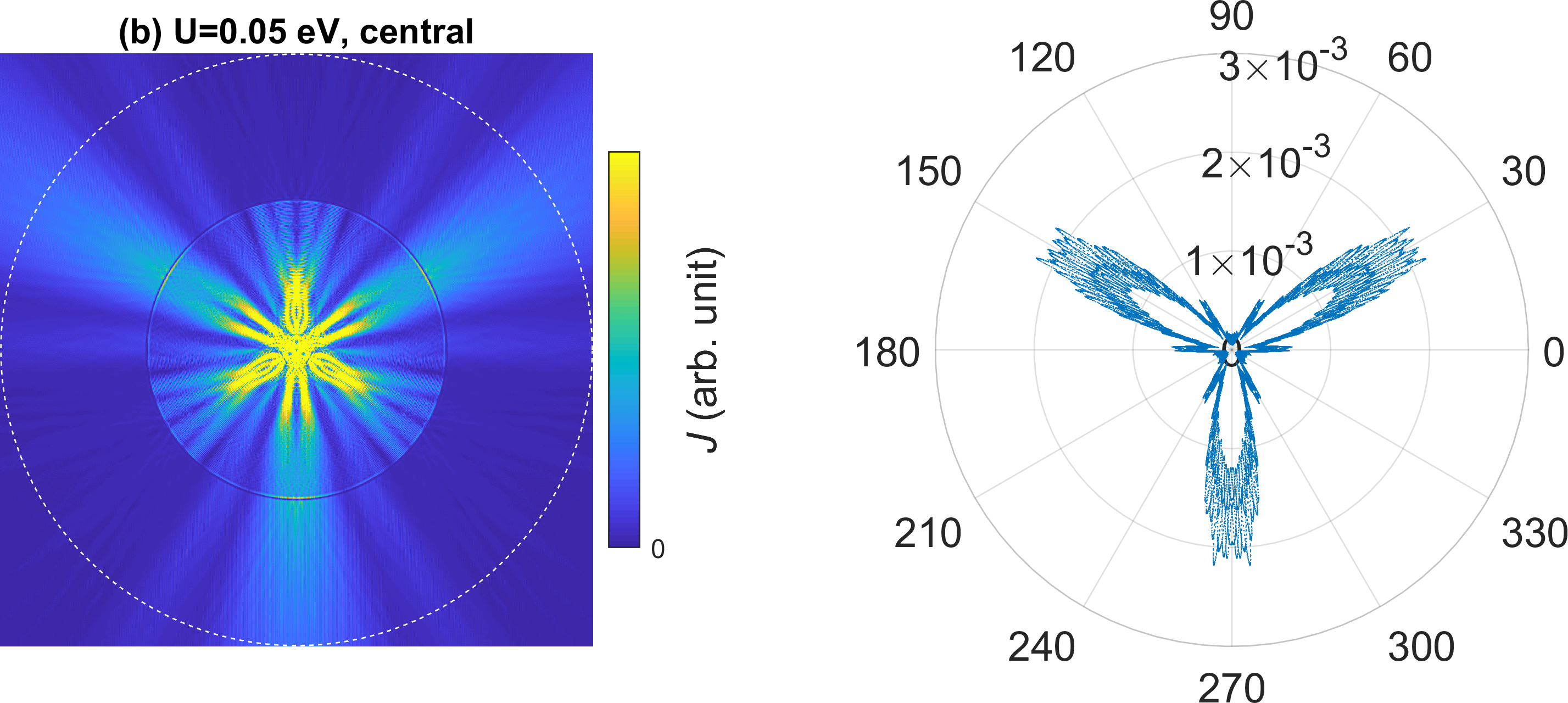}}
  \subfigure[]{\includegraphics[width=0.4\textwidth]{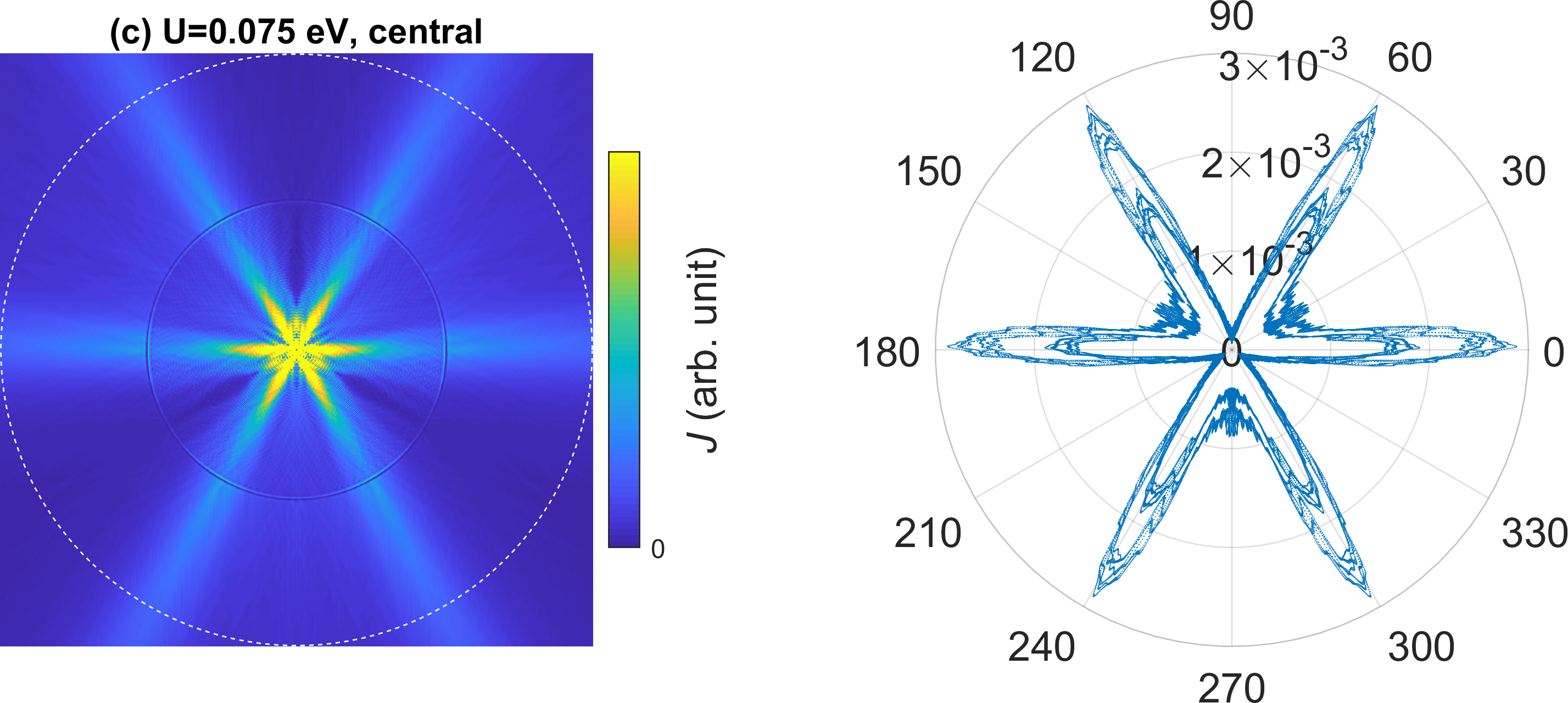}}
  \subfigure[]{\includegraphics[width=0.4\textwidth]{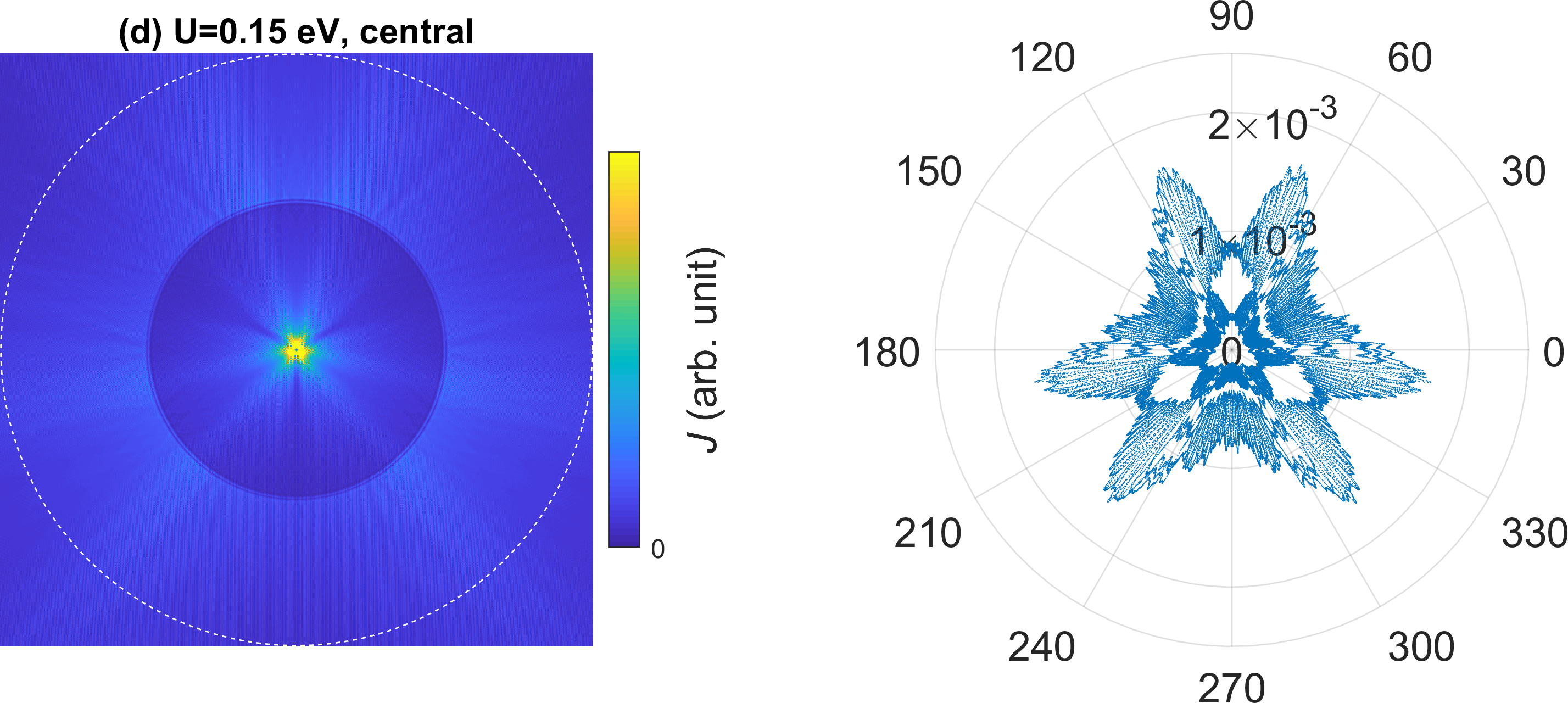}}
 \caption{BLG disk billiards ($\epsilon=0$) with source at central position $(x_s,y_s)=(0,0)$ and $V=0.1$ eV for different asymmetry parameters $U$. (a) $U= 0$ eV (anti-Klein tunneling) (b) $U= 0.05$ eV, (c) $U=0.075$ eV, and (d) $U=0.15$ eV modelled with full wave calculations. Left panels: local current density $J$ in real space, right panels: mid field emission polar plot ($r_m=2$). 
 Note that a trajectory-modelled source would not at all emit in (a) and would emit isotropically in (b,c,d), where wave simulations reveal the underlying hexagonal symmetry. 
 Certain parameters $U$ allow for efficient focusing such as in (c). 
  }
 \label{fig:bilayer_diskcentral}
\end{figure}

We now turn to BLG systems and start by investigating the classical-wave correspondence, cf.~Fig.~\ref{fig:raywavebilayer}. 
Typical resonance patterns and their polar emission plots are shown in Figs.~\ref{fig:raywavebilayer}(a,b) on the left and right, respectively. The formation of a whispering-gallery type mode is evident, in agreement with the importance of long trajectories in the simulated intensity: the anti-Klein tunneling feature of BLG allows the electrons emitted from the central source to remain inside the cavity despite their nearly normal incidence onto the boundary. Their subsequent dynamics in a cavity with nonlinear (chaotic) dynamics eventually allows the formation of whispering-gallery type modes that coexist with the source emission. Therefore, the emission pattern is a superposition of the long path contributions that can be associated with the Fresnel-weighted unstable manifold of the system (see the discussion in Sec.~\ref{sec:poincare} below), and short path contributions associated with the presence of a source. We will investigate and deepen the interesting features arising here in a separate work. 

In Fig.~\ref{fig:raywavebilayer}(c) the trajectory (wave) mid field emission $I$ ($J$) is shown for different source positions $(x_s, y_s=0)$. A reasonable agreement between classical and wave simulations is found.  
For central peak positions such as $x_s=0.4$, the simulations reveal emission peaks towards $\phi_m =0^o$ and $180^o$ in agreement with Fig.~\ref{fig:raywavebilayer}(a).

We point out the importance of short trajectories in the case of  near-boundary source positions, Fig.~\ref{fig:raywavebilayer}(a). It is those orbits that form of the strong emission peak into $\phi_m = 0^o$ (and, not shown, into direction $\phi_m=180^o$ for a source at the opposite boundary).
While this behaviour is reminiscent of SLG, long trajectories dominate otherwise as argued above. 

\subsection{Emission characteristics: Role of Fresnel laws}

\begin{figure}
\centering
 \subfigure[]{\includegraphics[width=0.44\textwidth]{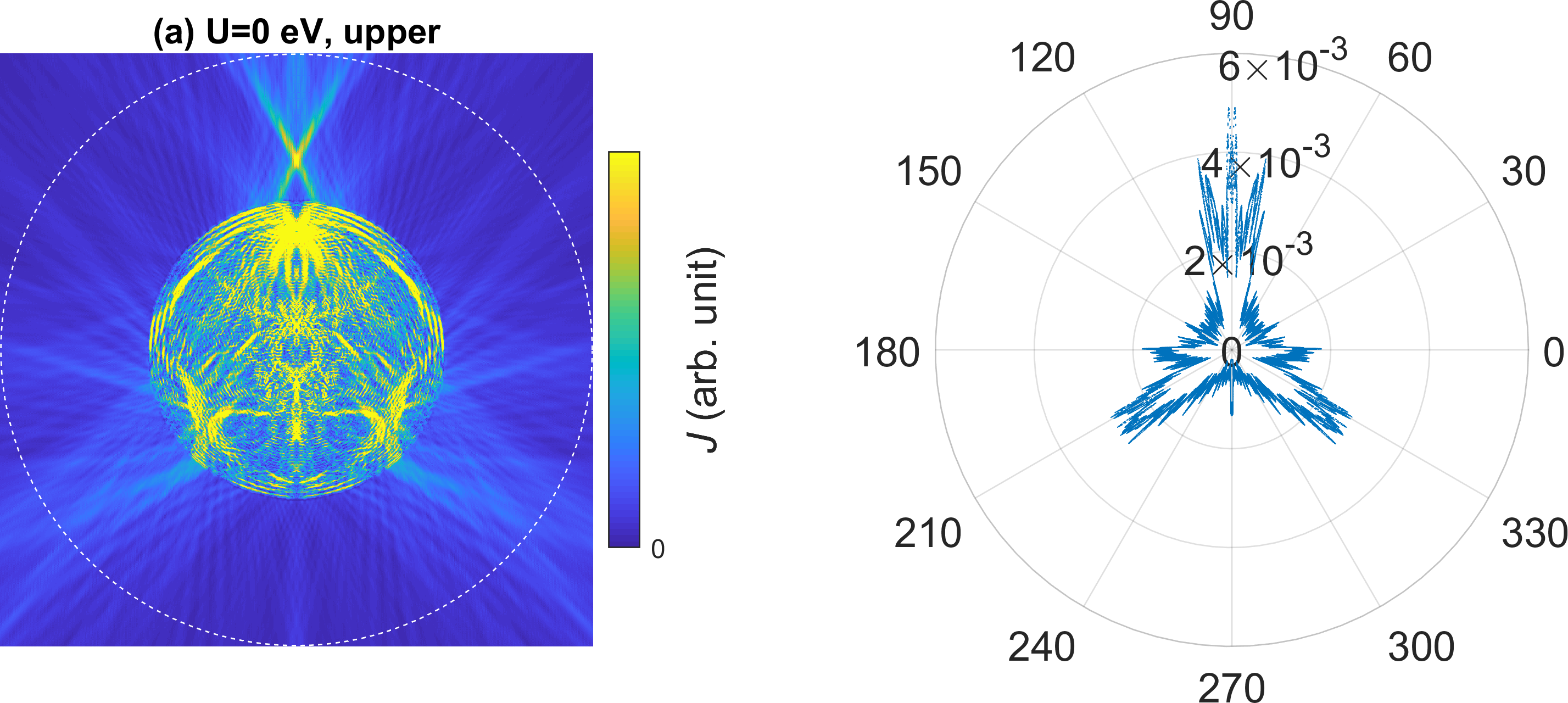}}
 \subfigure[]{\includegraphics[width=0.44\textwidth]{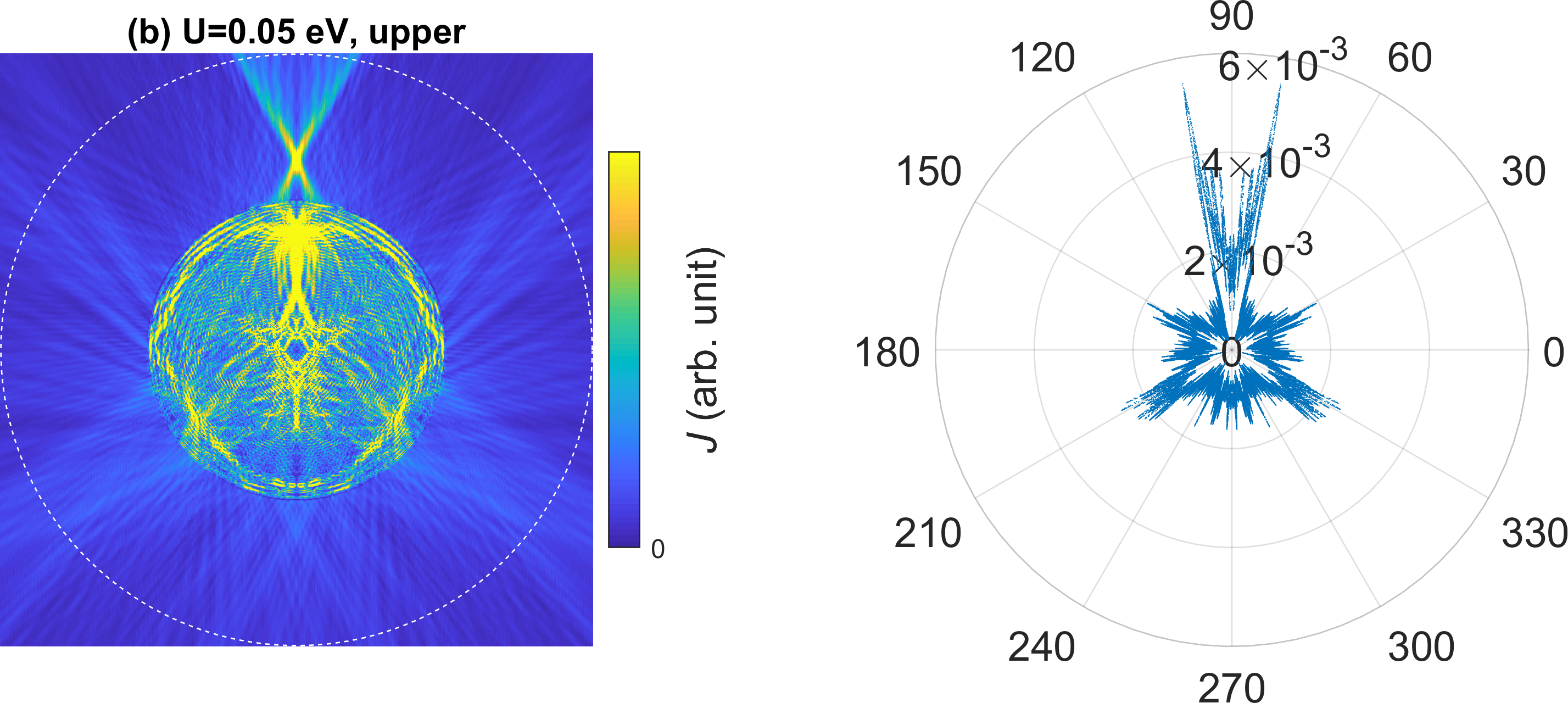}}
 \subfigure[]{\includegraphics[width=0.44\textwidth]{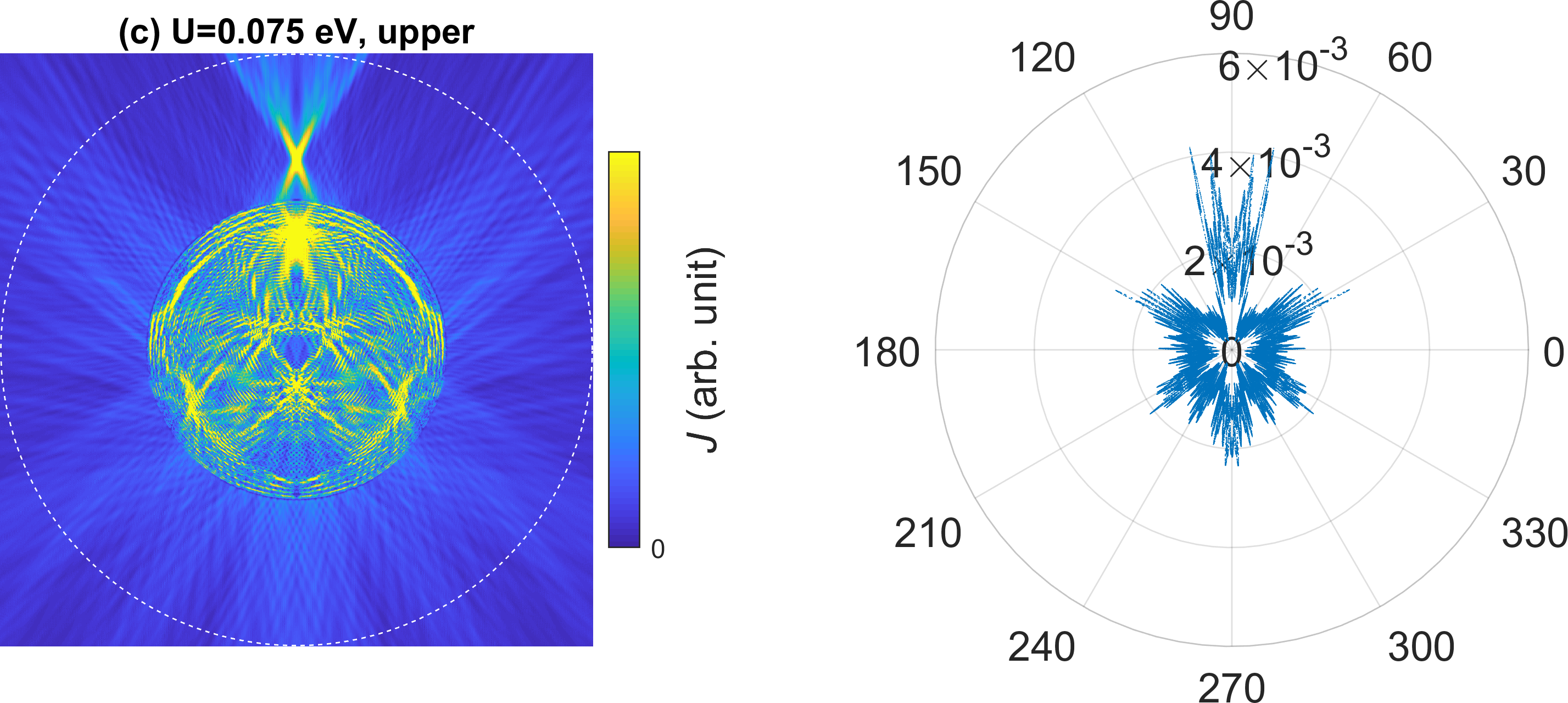}}
 \subfigure[]{\includegraphics[width=0.44\textwidth]{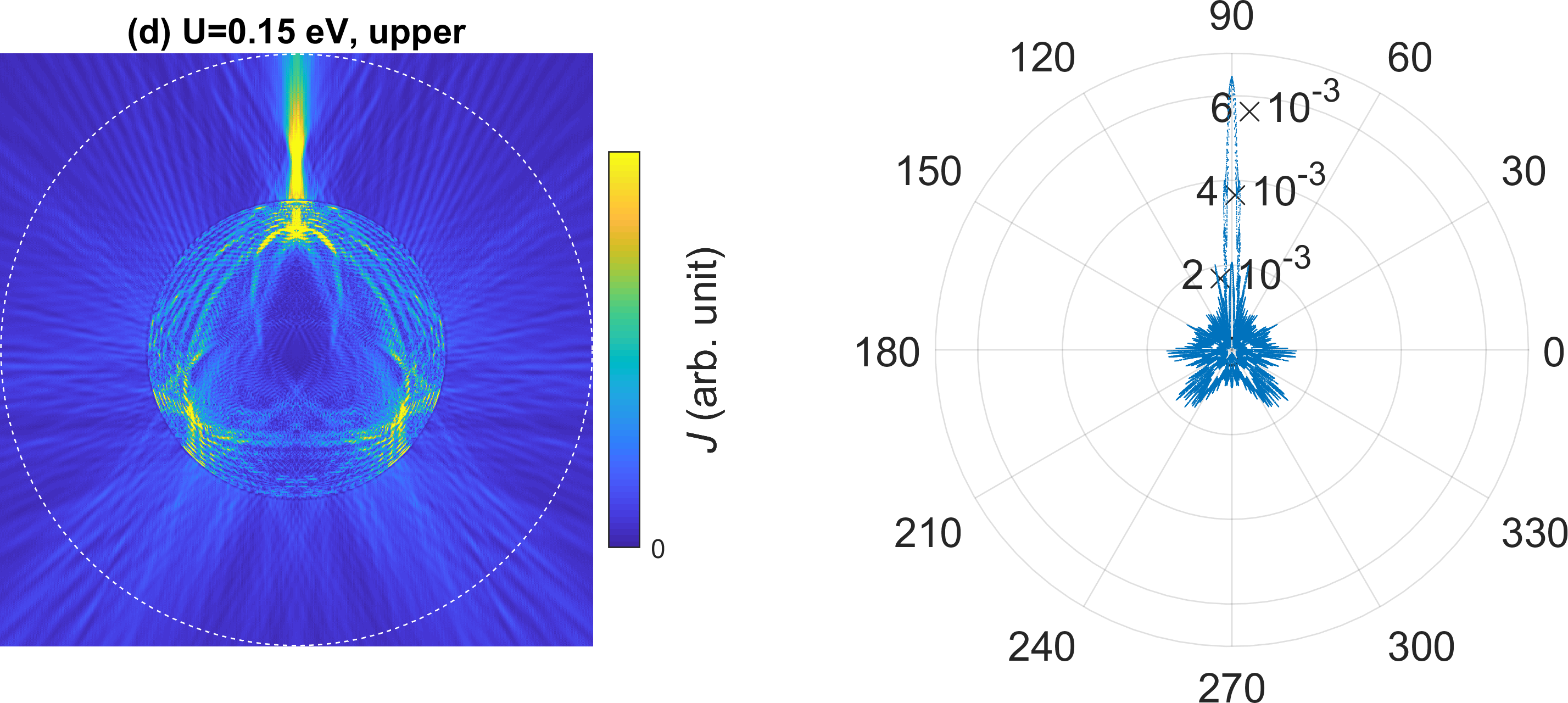}}
 \caption{Same as Fig.~\ref{fig:bilayer_diskcentral} 
 but with the source placed near the upper boundary at $(x_s,y_s)=(0,0.8)$. 
 Note the crossing beams that emerge in (a,b,c) near the source. They can be related to anti-Klein tunneling $T(\alpha=0)=0$ flanked by a maximumn of the Fresnel transmission coefficient for intermediate angles of incidence $\alpha$. The effect is lost when the regime of Klein tunneling is reached in (d). 
  }
 \label{fig:bilayer_diskupper}
\end{figure}

While we have illustrated the intricate interplay between the electron's dynamics and the Fresnel law for the orbit-based emission characteristics in various examples, we will now directly compare SLG and BLG systems.
Both possess very different Fresnel laws, cf.~Fig.~\ref{fig:fresnel}, and we will focus on how this affects their emission patterns using trajectory  simulations; cf.~Fig.~\ref{fig:singlelayerbilayer}.

\begin{figure}
\centering
 \subfigure[]{\includegraphics[width=0.43\textwidth]{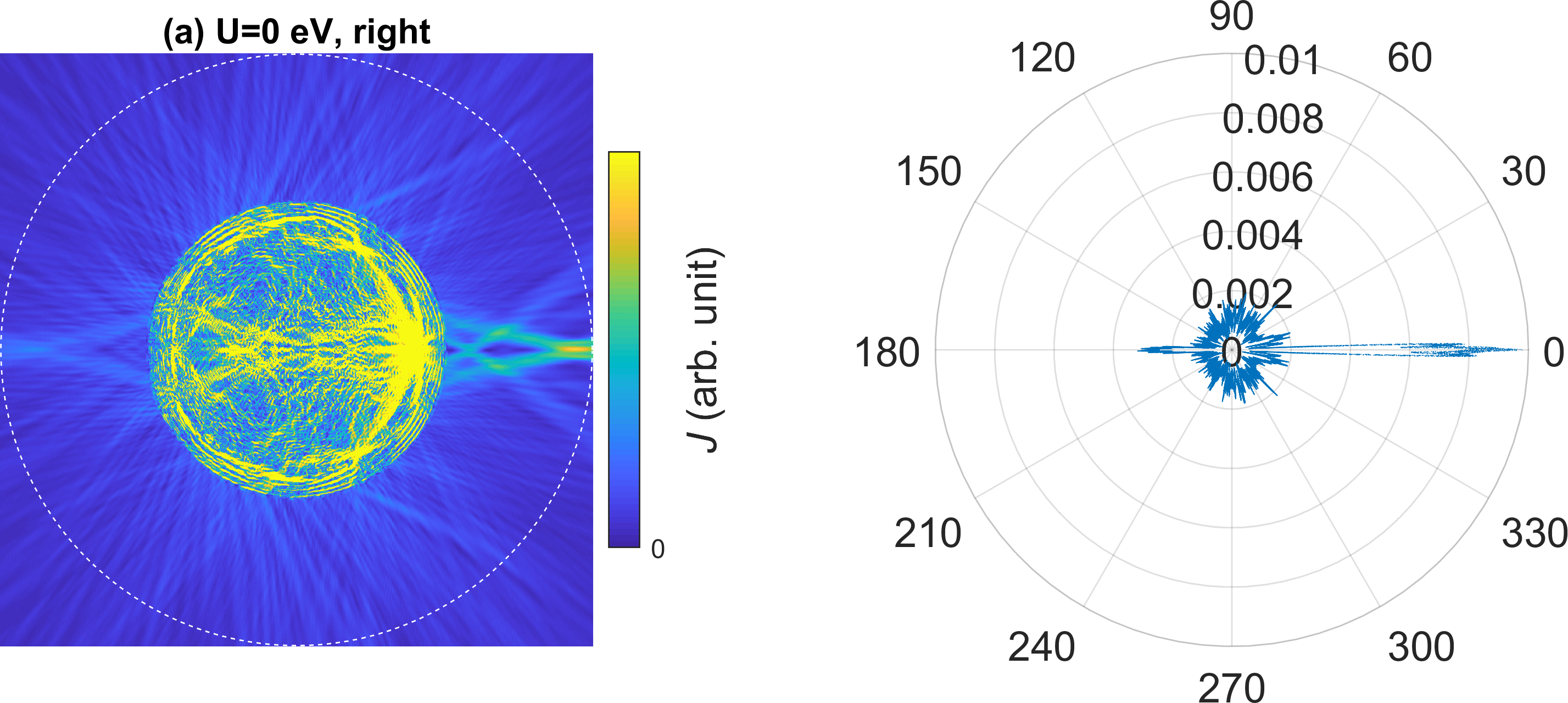}}
 \subfigure[]{\includegraphics[width=0.43\textwidth]{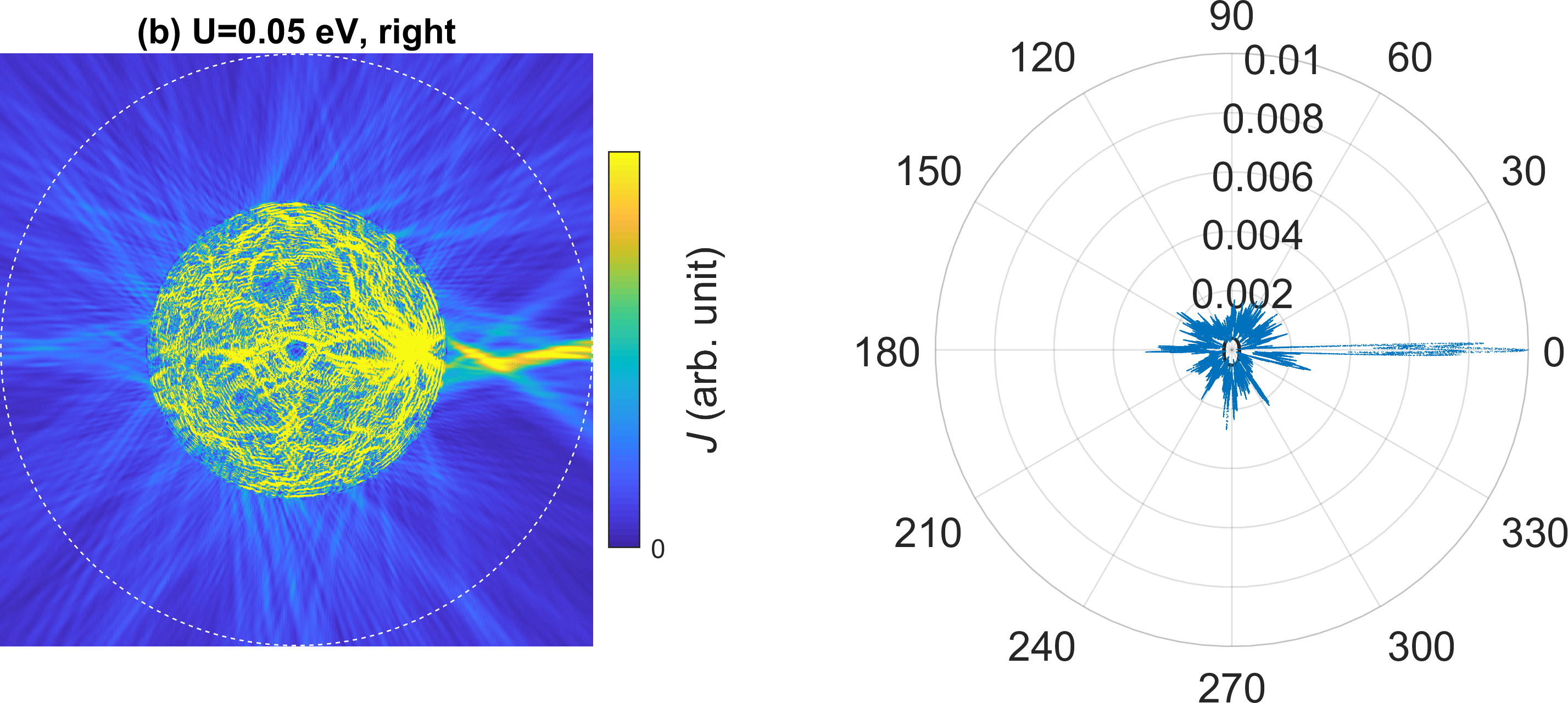}}
 \subfigure[]{\includegraphics[width=0.43\textwidth]{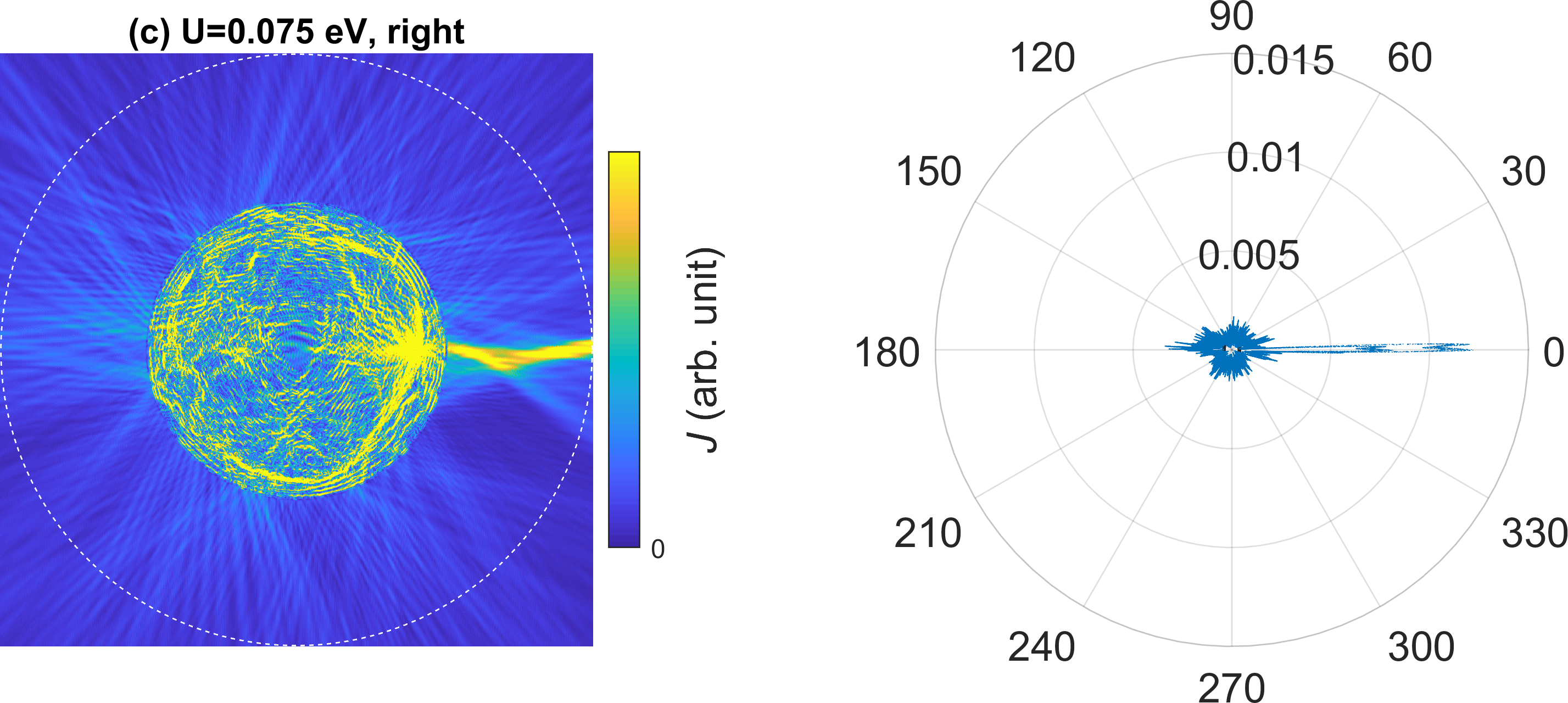}}
 \subfigure[]{\includegraphics[width=0.43\textwidth]{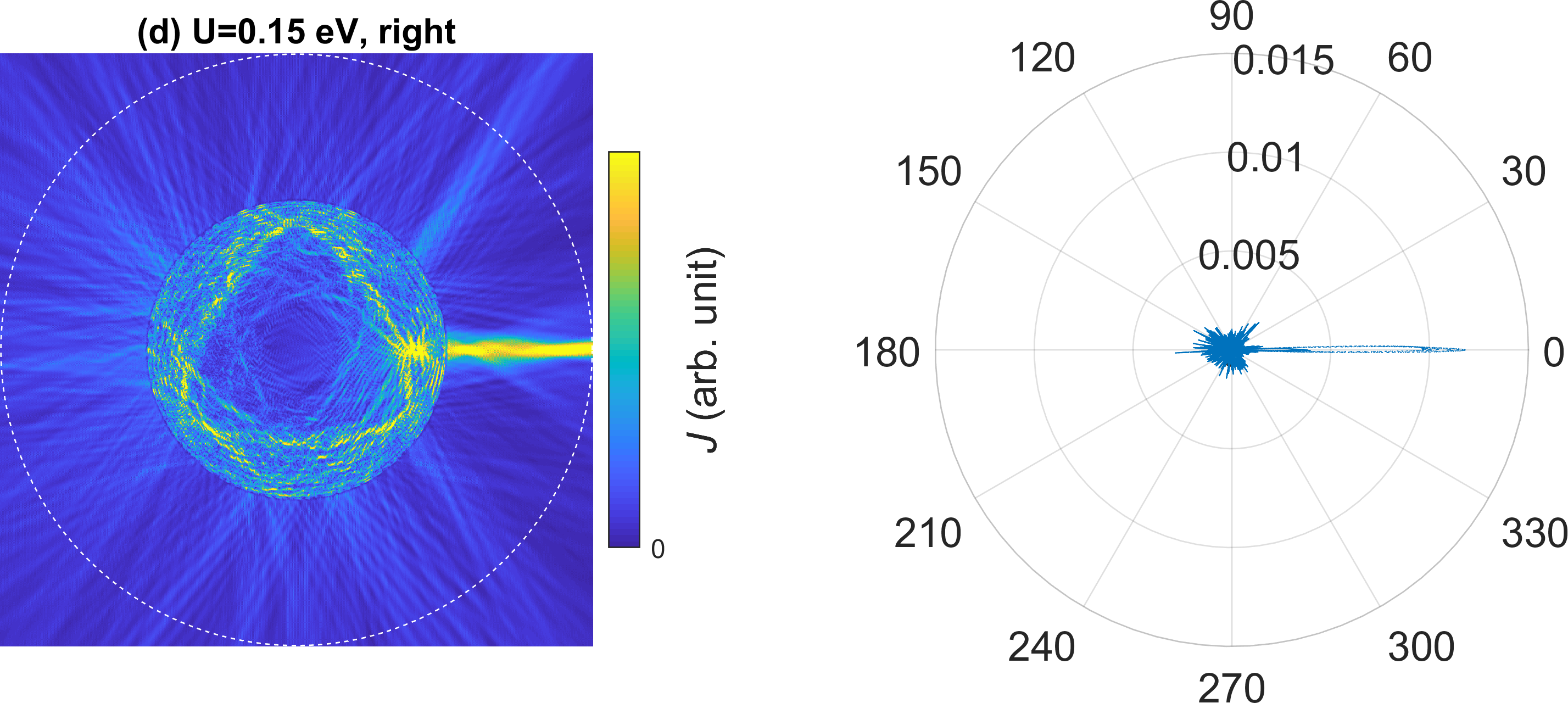}}
 \caption{Same as Fig.~\ref{fig:bilayer_diskcentral} 
 but with the source placed near the right boundary at $(x_s,y_s)=(0,0.8)$. 
 Although the geometry is symmetric with respect to the $x_s=x$-axis, 
 the intensity distribution $J$ is not since the $x$-axis is not a symmetry axis of the underlying BLG lattice. This yields deviations of the emission pattern in comparison to Fig.~\ref{fig:bilayer_diskupper} although the characteristic qualitative features -- main emission originating near the source, emergence of specific patterns -- are comparable.
  }
 \label{fig:bilayer_diskright}
\end{figure}

BLG has the interesting and remarkable property to perfectly reflect rays with normal incidence to the p-n interface and, thus, keep them inside the billiards. Consequently, the lensing effect disappears, and we expect the ray (and consequently the electron wave function or current density) intensity to become 
more independent from the source position. 
This is illustrated in Fig.~\ref{fig:singlelayerbilayer} for BLG in comparison to SLG and the optical case. To this end we compare the contribution of short trajectories, see Fig.~\ref{fig:singlelayerbilayer}(a) and long trajectories, see Fig.~\ref{fig:singlelayerbilayer}(b), for SLG  (green line) and BLG (blue and orange lines for effective refractive indices $n=-1$ and $n=-3$, respectively) systems. 

We begin our discussion with short trajectories, Fig.~\ref{fig:singlelayerbilayer}(a), where we recover 
the lensing effect for SLG 
(green line). 
As expected, the lensing effect disappears for BLG 
(blue and orange lines). At the same time, the importance of longer ray trajectories for bilayer systems is visible in Fig.~\ref{fig:singlelayerbilayer}(b), while long trajectories contribute very little to the mid field intensity for SLG (green line). Moreover, for BLG we find that the main emission directions depend on the refractive index $n$ chosen. This is directly related to the changes in the Fresnel coefficients, cf. Fig.~\ref{fig:fresnel}.

The optical case (TE polarisation, Fig.~\ref{fig:singlelayerbilayer}(c)),
is reminiscent of the behaviour of BLG. 
However, the optical mid field emission drops for central source positions around $x_s=0.5$, in stark contrast to the BLG case. The reason is the considerably better confinement for central source positions for BLG, 
resulting in normal ray incidence at the first reflection point. BLG yields perfect reflection in that case, while the optical reflection coefficient is smaller than one. 
For near-boundary optical sources, 
we observe the expected universal, source-position independent emission characteristics described in \cite{reviewletters2008}.


\subsection{Bilayer graphene disks: beyond particle-wave correspondence}
\label{sec_BLGbandstruct}

We have demonstrated the usefulness of a trajectory-based approach to graphene billiards. However, we will now illustrate the limits of particle-wave correspondence to be expected in a system of relativistic electrons on a honeycomb lattice. To this end we choose the example of circular (disk-like) BLG billiards (deformation parameter $\epsilon=0$). We will contrast a central source position that respects the rotational symmetry of the system, and source positions near the boundary.
In all cases we vary the asymmetry parameter $U$ \cite{McCannReview2013} such that we can freely scan from anti-Klein-tunneling for $U=0$ to situations between anti-Klein and Klein tunneling for $U>0$. We shall find an interpretation in terms of the Fresnel coefficient (in particular those for normal incidence) still useful, but at the same time other effects to become important. 

Figures \ref{fig:bilayer_diskcentral}, \ref{fig:bilayer_diskupper}, \ref{fig:bilayer_diskright} display a manifold of internal intensity distributions and mid field emission patterns as source position and asymmetry parameter $U$ are varied. We begin our discussion with a central source position, Fig.~\ref{fig:bilayer_diskcentral}. For $U=0$, Fig.~\ref{fig:bilayer_diskcentral}(a), we are in the regime of anti-Klein tunneling. Therefore, a trajectory simulation would show no emission into the mid field at all as all orbits emerging from the source will hit the boundary under normal incidence, i.e., a vanishing transmission coefficient. In Fig.~\ref{fig:bilayer_diskcentral}(b,c,d), where anti-Klein tunneling is lifted, an isotropic trajectory emission is expected. However, the full wave simulation results 
display 
distinct emission directions. This can be understood by a scattering argument: The current is injected isotropically from the point source. The reflected waves (anti-Klein tunneling) return to the point injector which now acts like a scatterer. However, the scattered waves are not simply s-wave-like, because the Fermi contour becomes non-circular for finite energies in BLG.
Hexagonal contributions to the Fermi contour $E(\vec{k})$ implies six predominant velocity directions $\vec{v} = \vec{\Delta}_{\vec{k}}E(\vec{k})$ as function of the wave vector $\vec{k}$, see Ref.~\cite{KraftPRL2020}.
This allows for the specific realization of patterns with highly focused emission directionality in few directions such a in  Fig.~\ref{fig:bilayer_diskcentral}(c) even if the geometry is literally concentric.

The situation changes when the source is placed near the system boundary as in Figs.~\ref{fig:bilayer_diskupper} and \ref{fig:bilayer_diskright}. For all $U$ considered, the main emission direction is related to the source position and pointing away from the source originating at the boundary position closest to the source. For a source placed near the upper boundary, Fig.~\ref{fig:bilayer_diskupper}, the resulting pattern is fully symmetric as is the underlying electronic BLG lattice. The electronic intensity is distributed over the whole disk, except for the case of Fig.~\ref{fig:bilayer_diskupper}(d) where a triangular pattern is formed. The mid field emission directions possess peaks in specific directions that are, however, different from those found for a central source position in Fig.~\ref{fig:bilayer_diskcentral}. 

A striking feature are the crossing beams leaving upwards in Fig.~\ref{fig:bilayer_diskupper}(a,b,c). Rays leaving the source will hit the boundary under different angles of incidence. For normal incidence there is no transmission due to anti-Klein tunneling. For increasing angles of incidence, the Fresnel transmission coefficient develops a maximum, cf.~Fig.~\ref{fig:fresnel}. This explains this crossing beam feature for 
$n=-1$. In Fig.~\ref{fig:bilayer_diskupper}(d), the maximum transmission is at normal incidence resulting in a single emerging beam. 

Finally, we discuss a source placed near the right boundary, Fig.~\ref{fig:bilayer_diskright}. The interplay of Fresnel-induced features with those originating in the electronic BLG lattice becomes particularly evident here. First of all, the geometric (shape) symmetry is not reflected in the intensity and mid field pattern, although the generic features discussed in Fig.~\ref{fig:bilayer_diskupper} can still be recognized. This nicely illustrates the limits of a ray interpretation and  ray-wave-correspondence in BLG systems and needs further consideration in another work.

\begin{figure}
\begin{center}
  \subfigure[]{\includegraphics[width=0.45\textwidth]{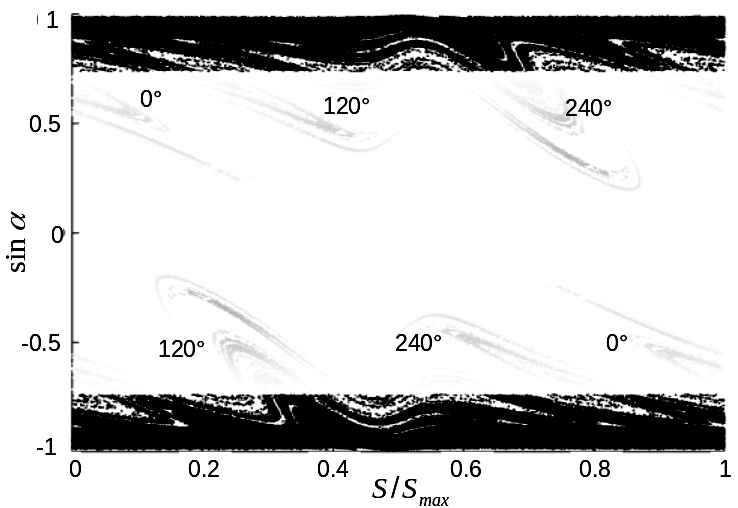}}
  \subfigure[]{\includegraphics[width=0.34\textwidth]{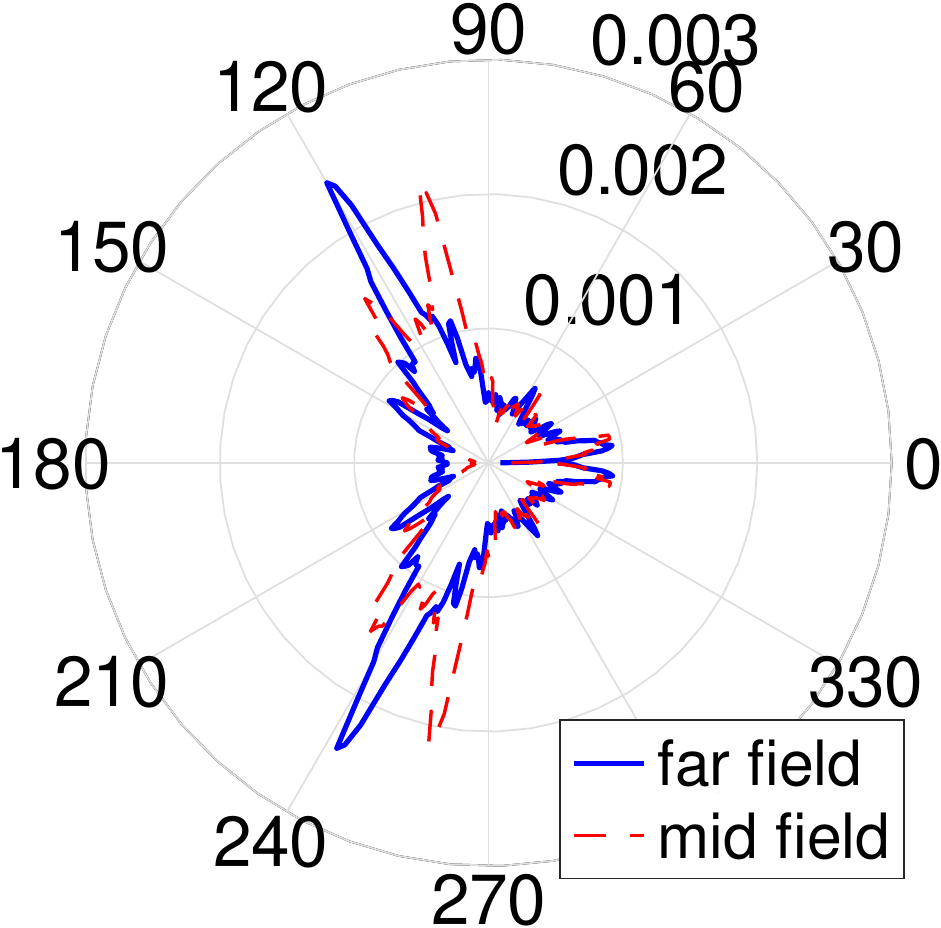}}
\end{center}
 \caption{(a) Trajectory-intensity weighted Poincar\'{e} surface of section for long orbits ($l>20$) in SLG with $n=-1$. Short paths are omitted in order to emphasize the filamentary structure of the natural measure originating from the unstable manifold of the ray dynamics. The filaments correspond to far field emission as indicated.
 (b) Farfield (continuous lines) and mid field (dotted lines) emission. One emission peak is caused by two peaks in the Poincar\'{e} map.} 
 \label{fig:poincare}
\end{figure}


\section{Conclusion and summary}
\label{sec:concl_summary}
\subsection{Emission characteristics: Role of carrier dynamics and Poincar\'{e} surface of section}
\label{sec:poincare}

We end our discussions with a trajectory-based analysis of our findings before the summary.
All deformed systems under investigation had the same geometry -- a lima\c{c}on with deformation parameter $\epsilon=0.43$. Therefore, the classical nonlinear, chaotic 
dynamics is the same in all (graphene and dielectric) systems considered here. This dynamics can be most efficiently captured in the so-called Poincar\'{e} surface of section (PSOS)\cite{stoecki_quantumchaos}. The PSOS 
is a mapping of trajectories from four-dimensional phase space to a two-dimensional space spanned by the so-called 
 Birkhoff coordinates, that is the arclength $s$ along the billiard's circumference of a reflection point and the sine of the corresponding angle of incidence, 
$\sin \alpha$; 
see Fig.~\ref{fig:poincare}(a). This representation comprises the location of the boundary reflection points and their tangential momentum for all trajectories and allows to visually access the system dynamics, e.g., to easily distinguish periodic (localized features in the PSOS) from chaotic dynamics.


Since the PSOS is the same for all systems considered here, all differences observed have to originate in the different, system-specific Fresnel laws. 
To include these into the PSOS, 
we weight the rayw by the respective Fresnel coefficient $R(\alpha)$ 
at this reflection point to yield their intensity $I$. 
This way different regions in the PSOS 
become differently important depending on the underlying material system's Fresnel law. The existence of a critical angle for total internal reflection in optical billiards, the perfect reflection at normal incidence for BLG, 
as well as the lesser confinement in SLG 
become visible in a ray-intensity weighted Poincar\'{e} surface of section;
Fig.~\ref{fig:poincare}(a) for SLG ($n=-1$, long rays). 
The fading intensity towards normal incidence ($\sin \alpha = 0$) corresponds to the increasing transmission coefficient (lesser ray confinement) for small angles of incidence $\alpha$. In turn, regions with grazing incidence have higher ray intensities. The system's emission characteristics, cf.~Fig.\ref{fig:poincare}(b), is determined by the transmitted intensity, i.e. by the filaments in the transmitting regions of the PSOS. 
Note that these filaments represent the unstable manifold of the system that will be populated by long ray trajectories \cite{reviewletters2008}. Each filament in the PSOS 
can be assigned a corresponding emission peak. Their far field emission direction is indicated by the angles given in Fig.~\ref{fig:poincare}(a). 

These considerations explain the specific emission patterns 
for SLG, BLG, and optical billiards  straightforwardly. 
It also reveals the expected slight deviation in the mid and far field directions and the better collimation of long rays in the far field. For other systems, the population of the Fresnel-weighted Poincer\'{e} surface of section will be different and reflect the underlying Fresnel law, and so will be the resulting emission characteristics. 



\subsection{Summary}
\label{sec:summary}

To summarize, we have studied the emission characteristics of single and bilayer graphene billiards with internal sources based on dynamical modelling that we justified by particle (ray)-wave correspondence and the use of appropriate Fresnel laws that were derived from exact wave considerations. 
We have outlined an experimental setup where our findings could be verified.

We found that the presence of sources and their position within the system may change the trajectories that dominate the emission into the mid/ far field from long trajectories that cover the description of optical microcavity lasers and bilayer-graphene billiards to short trajectories for single-layer graphene with central source positions. Which paths are of relevance depends on the structure of the Fresnel law (angular dependence of the transmission coefficient) in interplay with the Poincar\'{e} surface of section representing the system geometry. 
The dominance of short paths for single-layer graphene billiards with internal sources on the symmetry axis enables the manipulation of the mid field emission characteristics by a lensing effect made possible by the specific refractive index  $n=-1$ in single-layer graphene. By adjusting the source position, a strong focal-type collimation of ray intensity is realized in the mid field and semi-quantitatively confirmed by wave calculations. This effect can be further enhanced and manipulated by precisely engineering the boundary curvature of the billiard beyond the simple lima\c{c}on shape considered here.

\section{Acknowledgements}

K.R. acknowledges support from Deutsche Forschungsgemeinschaft (DFG, German Research Foundation), Project-IDs Ri681/13-1 and 314695032-SFB 1277 (project A07). 
 S.-C.C. and M.-H.L. are supported by Taiwan Ministry of Science and Technology (109-2112-M-006-020-MY3 and 107-2627-E-006-001).


\bibliography{bibtex_all}

\begin{thebibliography}{53}%
\makeatletter
\providecommand \@ifxundefined [1]{%
 \@ifx{#1\undefined}
}%
\providecommand \@ifnum [1]{%
 \ifnum #1\expandafter \@firstoftwo
 \else \expandafter \@secondoftwo
 \fi
}%
\providecommand \@ifx [1]{%
 \ifx #1\expandafter \@firstoftwo
 \else \expandafter \@secondoftwo
 \fi
}%
\providecommand \natexlab [1]{#1}%
\providecommand \enquote  [1]{``#1''}%
\providecommand \bibnamefont  [1]{#1}%
\providecommand \bibfnamefont [1]{#1}%
\providecommand \citenamefont [1]{#1}%
\providecommand \href@noop [0]{\@secondoftwo}%
\providecommand \href [0]{\begingroup \@sanitize@url \@href}%
\providecommand \@href[1]{\@@startlink{#1}\@@href}%
\providecommand \@@href[1]{\endgroup#1\@@endlink}%
\providecommand \@sanitize@url [0]{\catcode `\\12\catcode `\$12\catcode
  `\&12\catcode `\#12\catcode `\^12\catcode `\_12\catcode `\%12\relax}%
\providecommand \@@startlink[1]{}%
\providecommand \@@endlink[0]{}%
\providecommand \url  [0]{\begingroup\@sanitize@url \@url }%
\providecommand \@url [1]{\endgroup\@href {#1}{\urlprefix }}%
\providecommand \urlprefix  [0]{URL }%
\providecommand \Eprint [0]{\href }%
\providecommand \doibase [0]{http://dx.doi.org/}%
\providecommand \selectlanguage [0]{\@gobble}%
\providecommand \bibinfo  [0]{\@secondoftwo}%
\providecommand \bibfield  [0]{\@secondoftwo}%
\providecommand \translation [1]{[#1]}%
\providecommand \BibitemOpen [0]{}%
\providecommand \bibitemStop [0]{}%
\providecommand \bibitemNoStop [0]{.\EOS\space}%
\providecommand \EOS [0]{\spacefactor3000\relax}%
\providecommand \BibitemShut  [1]{\csname bibitem#1\endcsname}%
\let\auto@bib@innerbib\@empty
\bibitem [{\citenamefont {Katsnelson}\ \emph {et~al.}(2006)\citenamefont
  {Katsnelson}, \citenamefont {Novoselov},\ and\ \citenamefont {Geim}}]{KNG06}%
  \BibitemOpen
  \bibfield  {author} {\bibinfo {author} {\bibfnamefont {M.~I.}\ \bibnamefont
  {Katsnelson}}, \bibinfo {author} {\bibfnamefont {K.~S.}\ \bibnamefont
  {Novoselov}}, \ and\ \bibinfo {author} {\bibfnamefont {A.~K.}\ \bibnamefont
  {Geim}},\ }\href {\doibase 10.1038/nphys384} {\bibfield  {journal} {\bibinfo
  {journal} {Nature Physics}\ }\textbf {\bibinfo {volume} {2}},\ \bibinfo
  {pages} {620} (\bibinfo {year} {2006})}\BibitemShut {NoStop}%
\bibitem [{\citenamefont {Shytov}\ \emph {et~al.}(2008)\citenamefont {Shytov},
  \citenamefont {Rudner},\ and\ \citenamefont {Levitov}}]{Shytov2008}%
  \BibitemOpen
  \bibfield  {author} {\bibinfo {author} {\bibfnamefont {A.~V.}\ \bibnamefont
  {Shytov}}, \bibinfo {author} {\bibfnamefont {M.~S.}\ \bibnamefont {Rudner}},
  \ and\ \bibinfo {author} {\bibfnamefont {L.~S.}\ \bibnamefont {Levitov}},\
  }\href {\doibase 10.1103/physrevlett.101.156804} {\bibfield  {journal}
  {\bibinfo  {journal} {Physical Review Letters}\ }\textbf {\bibinfo {volume}
  {101}} (\bibinfo {year} {2008}),\ 10.1103/physrevlett.101.156804}\BibitemShut
  {NoStop}%
\bibitem [{\citenamefont {Young}\ and\ \citenamefont {Kim}(2009)}]{Young2009}%
  \BibitemOpen
  \bibfield  {author} {\bibinfo {author} {\bibfnamefont {A.~F.}\ \bibnamefont
  {Young}}\ and\ \bibinfo {author} {\bibfnamefont {P.}~\bibnamefont {Kim}},\
  }\href {\doibase 10.1038/nphys1198} {\bibfield  {journal} {\bibinfo
  {journal} {Nature Physics}\ }\textbf {\bibinfo {volume} {5}},\ \bibinfo
  {pages} {222} (\bibinfo {year} {2009})}\BibitemShut {NoStop}%
\bibitem [{\citenamefont {Masir}\ \emph {et~al.}(2010)\citenamefont {Masir},
  \citenamefont {Vasilopoulos},\ and\ \citenamefont {Peeters}}]{Masir2010}%
  \BibitemOpen
  \bibfield  {author} {\bibinfo {author} {\bibfnamefont {M.~R.}\ \bibnamefont
  {Masir}}, \bibinfo {author} {\bibfnamefont {P.}~\bibnamefont {Vasilopoulos}},
  \ and\ \bibinfo {author} {\bibfnamefont {F.~M.}\ \bibnamefont {Peeters}},\
  }\href {\doibase 10.1103/physrevb.82.115417} {\bibfield  {journal} {\bibinfo
  {journal} {Physical Review B}\ }\textbf {\bibinfo {volume} {82}} (\bibinfo
  {year} {2010}),\ 10.1103/physrevb.82.115417}\BibitemShut {NoStop}%
\bibitem [{\citenamefont {Nam}\ \emph {et~al.}(2011)\citenamefont {Nam},
  \citenamefont {Ki}, \citenamefont {Park}, \citenamefont {Kim}, \citenamefont
  {Kim},\ and\ \citenamefont {Lee}}]{Nam2011}%
  \BibitemOpen
  \bibfield  {author} {\bibinfo {author} {\bibfnamefont {S.-G.}\ \bibnamefont
  {Nam}}, \bibinfo {author} {\bibfnamefont {D.-K.}\ \bibnamefont {Ki}},
  \bibinfo {author} {\bibfnamefont {J.~W.}\ \bibnamefont {Park}}, \bibinfo
  {author} {\bibfnamefont {Y.}~\bibnamefont {Kim}}, \bibinfo {author}
  {\bibfnamefont {J.~S.}\ \bibnamefont {Kim}}, \ and\ \bibinfo {author}
  {\bibfnamefont {H.-J.}\ \bibnamefont {Lee}},\ }\href {\doibase
  10.1088/0957-4484/22/41/415203} {\bibfield  {journal} {\bibinfo  {journal}
  {Nanotechnology}\ }\textbf {\bibinfo {volume} {22}},\ \bibinfo {pages}
  {415203} (\bibinfo {year} {2011})}\BibitemShut {NoStop}%
\bibitem [{\citenamefont {Wang}\ \emph {et~al.}(2015)\citenamefont {Wang},
  \citenamefont {Yang}, \citenamefont {Chen}, \citenamefont {Watanabe},
  \citenamefont {Taniguchi}, \citenamefont {Churchill},\ and\ \citenamefont
  {Jarillo-Herrero}}]{Wang2015}%
  \BibitemOpen
  \bibfield  {author} {\bibinfo {author} {\bibfnamefont {J.~I.-J.}\
  \bibnamefont {Wang}}, \bibinfo {author} {\bibfnamefont {Y.}~\bibnamefont
  {Yang}}, \bibinfo {author} {\bibfnamefont {Y.-A.}\ \bibnamefont {Chen}},
  \bibinfo {author} {\bibfnamefont {K.}~\bibnamefont {Watanabe}}, \bibinfo
  {author} {\bibfnamefont {T.}~\bibnamefont {Taniguchi}}, \bibinfo {author}
  {\bibfnamefont {H.~O.~H.}\ \bibnamefont {Churchill}}, \ and\ \bibinfo
  {author} {\bibfnamefont {P.}~\bibnamefont {Jarillo-Herrero}},\ }\href
  {\doibase 10.1021/nl504750f} {\bibfield  {journal} {\bibinfo  {journal} {Nano
  Letters}\ }\textbf {\bibinfo {volume} {15}},\ \bibinfo {pages} {1898}
  (\bibinfo {year} {2015})}\BibitemShut {NoStop}%
\bibitem [{\citenamefont {Handschin}\ \emph {et~al.}(2016)\citenamefont
  {Handschin}, \citenamefont {Makk}, \citenamefont {Rickhaus}, \citenamefont
  {Liu}, \citenamefont {Watanabe}, \citenamefont {Taniguchi}, \citenamefont
  {Richter},\ and\ \citenamefont {Sch\"{o}nenberger}}]{Handschin2016}%
  \BibitemOpen
  \bibfield  {author} {\bibinfo {author} {\bibfnamefont {C.}~\bibnamefont
  {Handschin}}, \bibinfo {author} {\bibfnamefont {P.}~\bibnamefont {Makk}},
  \bibinfo {author} {\bibfnamefont {P.}~\bibnamefont {Rickhaus}}, \bibinfo
  {author} {\bibfnamefont {M.-H.}\ \bibnamefont {Liu}}, \bibinfo {author}
  {\bibfnamefont {K.}~\bibnamefont {Watanabe}}, \bibinfo {author}
  {\bibfnamefont {T.}~\bibnamefont {Taniguchi}}, \bibinfo {author}
  {\bibfnamefont {K.}~\bibnamefont {Richter}}, \ and\ \bibinfo {author}
  {\bibfnamefont {C.}~\bibnamefont {Sch\"{o}nenberger}},\ }\href {\doibase
  10.1021/acs.nanolett.6b04137} {\bibfield  {journal} {\bibinfo  {journal}
  {Nano Letters}\ }\textbf {\bibinfo {volume} {17}},\ \bibinfo {pages} {328}
  (\bibinfo {year} {2016})}\BibitemShut {NoStop}%
\bibitem [{\citenamefont {Cheianov}\ and\ \citenamefont {Fal'ko}(2006)}]{CF06}%
  \BibitemOpen
  \bibfield  {author} {\bibinfo {author} {\bibfnamefont {V.~V.}\ \bibnamefont
  {Cheianov}}\ and\ \bibinfo {author} {\bibfnamefont {V.~I.}\ \bibnamefont
  {Fal'ko}},\ }\href {\doibase 10.1103/physrevb.74.041403} {\bibfield
  {journal} {\bibinfo  {journal} {Physical Review B}\ }\textbf {\bibinfo
  {volume} {74}} (\bibinfo {year} {2006}),\
  10.1103/physrevb.74.041403}\BibitemShut {NoStop}%
\bibitem [{\citenamefont {Rickhaus}\ \emph {et~al.}(2013)\citenamefont
  {Rickhaus}, \citenamefont {Maurand}, \citenamefont {Liu}, \citenamefont
  {Weiss}, \citenamefont {Richter},\ and\ \citenamefont
  {Sch\"{o}nenberger}}]{RML+13}%
  \BibitemOpen
  \bibfield  {author} {\bibinfo {author} {\bibfnamefont {P.}~\bibnamefont
  {Rickhaus}}, \bibinfo {author} {\bibfnamefont {R.}~\bibnamefont {Maurand}},
  \bibinfo {author} {\bibfnamefont {M.-H.}\ \bibnamefont {Liu}}, \bibinfo
  {author} {\bibfnamefont {M.}~\bibnamefont {Weiss}}, \bibinfo {author}
  {\bibfnamefont {K.}~\bibnamefont {Richter}}, \ and\ \bibinfo {author}
  {\bibfnamefont {C.}~\bibnamefont {Sch\"{o}nenberger}},\ }\href
  {https://epub.uni-regensburg.de/28731/} {\bibfield  {journal} {\bibinfo
  {journal} {Nature Communications}\ }\textbf {\bibinfo {volume} {4}},\
  \bibinfo {pages} {2342} (\bibinfo {year} {2013})}\BibitemShut {NoStop}%
\bibitem [{\citenamefont {Grushina}\ \emph {et~al.}(2013)\citenamefont
  {Grushina}, \citenamefont {Ki},\ and\ \citenamefont
  {Morpurgo}}]{Grushina2013}%
  \BibitemOpen
  \bibfield  {author} {\bibinfo {author} {\bibfnamefont {A.~L.}\ \bibnamefont
  {Grushina}}, \bibinfo {author} {\bibfnamefont {D.-K.}\ \bibnamefont {Ki}}, \
  and\ \bibinfo {author} {\bibfnamefont {A.~F.}\ \bibnamefont {Morpurgo}},\
  }\href {\doibase 10.1063/1.4807888} {\bibfield  {journal} {\bibinfo
  {journal} {Applied Physics Letters}\ }\textbf {\bibinfo {volume} {102}},\
  \bibinfo {pages} {223102} (\bibinfo {year} {2013})}\BibitemShut {NoStop}%
\bibitem [{\citenamefont {Kraft}\ \emph {et~al.}(2020)\citenamefont {Kraft},
  \citenamefont {Liu}, \citenamefont {Selvasundaram}, \citenamefont {Chen},
  \citenamefont {Krupke}, \citenamefont {Richter},\ and\ \citenamefont
  {Danneau}}]{KraftPRL2020}%
  \BibitemOpen
  \bibfield  {author} {\bibinfo {author} {\bibfnamefont {R.}~\bibnamefont
  {Kraft}}, \bibinfo {author} {\bibfnamefont {M.-H.}\ \bibnamefont {Liu}},
  \bibinfo {author} {\bibfnamefont {P.~B.}\ \bibnamefont {Selvasundaram}},
  \bibinfo {author} {\bibfnamefont {S.-C.}\ \bibnamefont {Chen}}, \bibinfo
  {author} {\bibfnamefont {R.}~\bibnamefont {Krupke}}, \bibinfo {author}
  {\bibfnamefont {K.}~\bibnamefont {Richter}}, \ and\ \bibinfo {author}
  {\bibfnamefont {R.}~\bibnamefont {Danneau}},\ }\href {\doibase
  10.1103/physrevlett.125.217701} {\bibfield  {journal} {\bibinfo  {journal}
  {Physical Review Letters}\ }\textbf {\bibinfo {volume} {125}} (\bibinfo
  {year} {2020}),\ 10.1103/physrevlett.125.217701}\BibitemShut {NoStop}%
\bibitem [{\citenamefont {Rehmann}\ \emph {et~al.}(2019)\citenamefont
  {Rehmann}, \citenamefont {Kalyoncu}, \citenamefont {Kisiel}, \citenamefont
  {Pascher}, \citenamefont {Giessibl}, \citenamefont {M\"{u}ller},
  \citenamefont {Watanabe}, \citenamefont {Taniguchi}, \citenamefont {Meyer},
  \citenamefont {Liu},\ and\ \citenamefont {Zumb\"{u}hl}}]{Rehmann2019}%
  \BibitemOpen
  \bibfield  {author} {\bibinfo {author} {\bibfnamefont {M.~K.}\ \bibnamefont
  {Rehmann}}, \bibinfo {author} {\bibfnamefont {Y.~B.}\ \bibnamefont
  {Kalyoncu}}, \bibinfo {author} {\bibfnamefont {M.}~\bibnamefont {Kisiel}},
  \bibinfo {author} {\bibfnamefont {N.}~\bibnamefont {Pascher}}, \bibinfo
  {author} {\bibfnamefont {F.~J.}\ \bibnamefont {Giessibl}}, \bibinfo {author}
  {\bibfnamefont {F.}~\bibnamefont {M\"{u}ller}}, \bibinfo {author}
  {\bibfnamefont {K.}~\bibnamefont {Watanabe}}, \bibinfo {author}
  {\bibfnamefont {T.}~\bibnamefont {Taniguchi}}, \bibinfo {author}
  {\bibfnamefont {E.}~\bibnamefont {Meyer}}, \bibinfo {author} {\bibfnamefont
  {M.-H.}\ \bibnamefont {Liu}}, \ and\ \bibinfo {author} {\bibfnamefont
  {D.~M.}\ \bibnamefont {Zumb\"{u}hl}},\ }\href {\doibase
  10.1016/j.carbon.2019.05.015} {\bibfield  {journal} {\bibinfo  {journal}
  {Carbon}\ }\textbf {\bibinfo {volume} {150}},\ \bibinfo {pages} {417}
  (\bibinfo {year} {2019})}\BibitemShut {NoStop}%
\bibitem [{\citenamefont {Varlet}\ \emph {et~al.}(2014)\citenamefont {Varlet},
  \citenamefont {Liu}, \citenamefont {Krueckl}, \citenamefont {Bischoff},
  \citenamefont {Simonet}, \citenamefont {Watanabe}, \citenamefont {Taniguchi},
  \citenamefont {Richter}, \citenamefont {Ensslin},\ and\ \citenamefont
  {Ihn}}]{VLK+14}%
  \BibitemOpen
  \bibfield  {author} {\bibinfo {author} {\bibfnamefont {A.}~\bibnamefont
  {Varlet}}, \bibinfo {author} {\bibfnamefont {M.-H.}\ \bibnamefont {Liu}},
  \bibinfo {author} {\bibfnamefont {V.}~\bibnamefont {Krueckl}}, \bibinfo
  {author} {\bibfnamefont {D.}~\bibnamefont {Bischoff}}, \bibinfo {author}
  {\bibfnamefont {P.}~\bibnamefont {Simonet}}, \bibinfo {author} {\bibfnamefont
  {K.}~\bibnamefont {Watanabe}}, \bibinfo {author} {\bibfnamefont
  {T.}~\bibnamefont {Taniguchi}}, \bibinfo {author} {\bibfnamefont
  {K.}~\bibnamefont {Richter}}, \bibinfo {author} {\bibfnamefont
  {K.}~\bibnamefont {Ensslin}}, \ and\ \bibinfo {author} {\bibfnamefont
  {T.}~\bibnamefont {Ihn}},\ }\href {\doibase 10.1103/physrevlett.113.116601}
  {\bibfield  {journal} {\bibinfo  {journal} {Physical Review Letters}\
  }\textbf {\bibinfo {volume} {113}} (\bibinfo {year} {2014}),\
  10.1103/physrevlett.113.116601}\BibitemShut {NoStop}%
\bibitem [{\citenamefont {Varlet}\ \emph {et~al.}(2015)\citenamefont {Varlet},
  \citenamefont {Liu}, \citenamefont {Bischoff}, \citenamefont {Simonet},
  \citenamefont {Taniguchi}, \citenamefont {Watanabe}, \citenamefont {Richter},
  \citenamefont {Ihn},\ and\ \citenamefont {Ensslin}}]{Varlet2015}%
  \BibitemOpen
  \bibfield  {author} {\bibinfo {author} {\bibfnamefont {A.}~\bibnamefont
  {Varlet}}, \bibinfo {author} {\bibfnamefont {M.-H.}\ \bibnamefont {Liu}},
  \bibinfo {author} {\bibfnamefont {D.}~\bibnamefont {Bischoff}}, \bibinfo
  {author} {\bibfnamefont {P.}~\bibnamefont {Simonet}}, \bibinfo {author}
  {\bibfnamefont {T.}~\bibnamefont {Taniguchi}}, \bibinfo {author}
  {\bibfnamefont {K.}~\bibnamefont {Watanabe}}, \bibinfo {author}
  {\bibfnamefont {K.}~\bibnamefont {Richter}}, \bibinfo {author} {\bibfnamefont
  {T.}~\bibnamefont {Ihn}}, \ and\ \bibinfo {author} {\bibfnamefont
  {K.}~\bibnamefont {Ensslin}},\ }\href {\doibase 10.1002/pssr.201510180}
  {\bibfield  {journal} {\bibinfo  {journal} {physica status solidi ({RRL}) -
  Rapid Research Letters}\ }\textbf {\bibinfo {volume} {10}},\ \bibinfo {pages}
  {46} (\bibinfo {year} {2015})}\BibitemShut {NoStop}%
\bibitem [{\citenamefont {Park}\ and\ \citenamefont {Sim}(2011)}]{Park2011}%
  \BibitemOpen
  \bibfield  {author} {\bibinfo {author} {\bibfnamefont {S.}~\bibnamefont
  {Park}}\ and\ \bibinfo {author} {\bibfnamefont {H.-S.}\ \bibnamefont {Sim}},\
  }\href {\doibase 10.1103/physrevb.84.235432} {\bibfield  {journal} {\bibinfo
  {journal} {Physical Review B}\ }\textbf {\bibinfo {volume} {84}} (\bibinfo
  {year} {2011}),\ 10.1103/physrevb.84.235432}\BibitemShut {NoStop}%
\bibitem [{\citenamefont {Du}\ \emph {et~al.}(2018)\citenamefont {Du},
  \citenamefont {Liu}, \citenamefont {Mohrmann}, \citenamefont {Wu},
  \citenamefont {Krupke}, \citenamefont {von L\"{o}hneysen}, \citenamefont
  {Richter},\ and\ \citenamefont {Danneau}}]{RLM+18}%
  \BibitemOpen
  \bibfield  {author} {\bibinfo {author} {\bibfnamefont {R.}~\bibnamefont
  {Du}}, \bibinfo {author} {\bibfnamefont {M.-H.}\ \bibnamefont {Liu}},
  \bibinfo {author} {\bibfnamefont {J.}~\bibnamefont {Mohrmann}}, \bibinfo
  {author} {\bibfnamefont {F.}~\bibnamefont {Wu}}, \bibinfo {author}
  {\bibfnamefont {R.}~\bibnamefont {Krupke}}, \bibinfo {author} {\bibfnamefont
  {H.}~\bibnamefont {von L\"{o}hneysen}}, \bibinfo {author} {\bibfnamefont
  {K.}~\bibnamefont {Richter}}, \ and\ \bibinfo {author} {\bibfnamefont
  {R.}~\bibnamefont {Danneau}},\ }\href {\doibase
  10.1103/physrevlett.121.127706} {\bibfield  {journal} {\bibinfo  {journal}
  {Physical Review Letters}\ }\textbf {\bibinfo {volume} {121}} (\bibinfo
  {year} {2018}),\ 10.1103/physrevlett.121.127706}\BibitemShut {NoStop}%
\bibitem [{\citenamefont {Rickhaus}\ \emph {et~al.}(2020)\citenamefont
  {Rickhaus}, \citenamefont {Liu}, \citenamefont {Kurpas}, \citenamefont
  {Kurzmann}, \citenamefont {Lee}, \citenamefont {Overweg}, \citenamefont
  {Eich}, \citenamefont {Pisoni}, \citenamefont {Taniguchi}, \citenamefont
  {Watanabe}, \citenamefont {Richter}, \citenamefont {Ensslin},\ and\
  \citenamefont {Ihn}}]{Rickhaus2020}%
  \BibitemOpen
  \bibfield  {author} {\bibinfo {author} {\bibfnamefont {P.}~\bibnamefont
  {Rickhaus}}, \bibinfo {author} {\bibfnamefont {M.-H.}\ \bibnamefont {Liu}},
  \bibinfo {author} {\bibfnamefont {M.}~\bibnamefont {Kurpas}}, \bibinfo
  {author} {\bibfnamefont {A.}~\bibnamefont {Kurzmann}}, \bibinfo {author}
  {\bibfnamefont {Y.}~\bibnamefont {Lee}}, \bibinfo {author} {\bibfnamefont
  {H.}~\bibnamefont {Overweg}}, \bibinfo {author} {\bibfnamefont
  {M.}~\bibnamefont {Eich}}, \bibinfo {author} {\bibfnamefont {R.}~\bibnamefont
  {Pisoni}}, \bibinfo {author} {\bibfnamefont {T.}~\bibnamefont {Taniguchi}},
  \bibinfo {author} {\bibfnamefont {K.}~\bibnamefont {Watanabe}}, \bibinfo
  {author} {\bibfnamefont {K.}~\bibnamefont {Richter}}, \bibinfo {author}
  {\bibfnamefont {K.}~\bibnamefont {Ensslin}}, \ and\ \bibinfo {author}
  {\bibfnamefont {T.}~\bibnamefont {Ihn}},\ }\href {\doibase
  10.1126/sciadv.aay8409} {\bibfield  {journal} {\bibinfo  {journal} {Science
  Advances}\ }\textbf {\bibinfo {volume} {6}},\ \bibinfo {pages} {eaay8409}
  (\bibinfo {year} {2020})}\BibitemShut {NoStop}%
\bibitem [{\citenamefont {P{\'{e}}terfalvi}\ \emph {et~al.}(2009)\citenamefont
  {P{\'{e}}terfalvi}, \citenamefont {P{\'{a}}lyi},\ and\ \citenamefont
  {Cserti}}]{Peterfalvi2009}%
  \BibitemOpen
  \bibfield  {author} {\bibinfo {author} {\bibfnamefont {C.}~\bibnamefont
  {P{\'{e}}terfalvi}}, \bibinfo {author} {\bibfnamefont {A.}~\bibnamefont
  {P{\'{a}}lyi}}, \ and\ \bibinfo {author} {\bibfnamefont {J.}~\bibnamefont
  {Cserti}},\ }\href {\doibase 10.1103/physrevb.80.075416} {\bibfield
  {journal} {\bibinfo  {journal} {Physical Review B}\ }\textbf {\bibinfo
  {volume} {80}} (\bibinfo {year} {2009}),\
  10.1103/physrevb.80.075416}\BibitemShut {NoStop}%
\bibitem [{\citenamefont {Wang}\ \emph {et~al.}(2019)\citenamefont {Wang},
  \citenamefont {Elahi}, \citenamefont {Wang}, \citenamefont {Habib},
  \citenamefont {Taniguchi}, \citenamefont {Watanabe}, \citenamefont {Hone},
  \citenamefont {Ghosh}, \citenamefont {Lee},\ and\ \citenamefont
  {Kim}}]{Wang19}%
  \BibitemOpen
  \bibfield  {author} {\bibinfo {author} {\bibfnamefont {K.}~\bibnamefont
  {Wang}}, \bibinfo {author} {\bibfnamefont {M.~M.}\ \bibnamefont {Elahi}},
  \bibinfo {author} {\bibfnamefont {L.}~\bibnamefont {Wang}}, \bibinfo {author}
  {\bibfnamefont {K.~M.~M.}\ \bibnamefont {Habib}}, \bibinfo {author}
  {\bibfnamefont {T.}~\bibnamefont {Taniguchi}}, \bibinfo {author}
  {\bibfnamefont {K.}~\bibnamefont {Watanabe}}, \bibinfo {author}
  {\bibfnamefont {J.}~\bibnamefont {Hone}}, \bibinfo {author} {\bibfnamefont
  {A.~W.}\ \bibnamefont {Ghosh}}, \bibinfo {author} {\bibfnamefont {G.-H.}\
  \bibnamefont {Lee}}, \ and\ \bibinfo {author} {\bibfnamefont
  {P.}~\bibnamefont {Kim}},\ }\href {\doibase 10.1073/pnas.1816119116}
  {\bibfield  {journal} {\bibinfo  {journal} {Proceedings of the National
  Academy of Sciences}\ }\textbf {\bibinfo {volume} {116}},\ \bibinfo {pages}
  {6575} (\bibinfo {year} {2019})}\BibitemShut {NoStop}%
\bibitem [{\citenamefont {Cheianov}\ \emph {et~al.}(2007)\citenamefont
  {Cheianov}, \citenamefont {Fal'ko},\ and\ \citenamefont {Altshuler}}]{CFA07}%
  \BibitemOpen
  \bibfield  {author} {\bibinfo {author} {\bibfnamefont {V.~V.}\ \bibnamefont
  {Cheianov}}, \bibinfo {author} {\bibfnamefont {V.}~\bibnamefont {Fal'ko}}, \
  and\ \bibinfo {author} {\bibfnamefont {B.~L.}\ \bibnamefont {Altshuler}},\
  }\href {\doibase 10.1126/science.1138020} {\bibfield  {journal} {\bibinfo
  {journal} {Science}\ }\textbf {\bibinfo {volume} {315}},\ \bibinfo {pages}
  {1252} (\bibinfo {year} {2007})}\BibitemShut {NoStop}%
\bibitem [{\citenamefont {Liu}\ \emph {et~al.}(2017)\citenamefont {Liu},
  \citenamefont {Gorini},\ and\ \citenamefont {Richter}}]{LGR17}%
  \BibitemOpen
  \bibfield  {author} {\bibinfo {author} {\bibfnamefont {M.-H.}\ \bibnamefont
  {Liu}}, \bibinfo {author} {\bibfnamefont {C.}~\bibnamefont {Gorini}}, \ and\
  \bibinfo {author} {\bibfnamefont {K.}~\bibnamefont {Richter}},\ }\href
  {\doibase 10.1103/PhysRevLett.118.066801} {\bibfield  {journal} {\bibinfo
  {journal} {Phys. Rev. Lett.}\ }\textbf {\bibinfo {volume} {118}},\ \bibinfo
  {pages} {066801} (\bibinfo {year} {2017})}\BibitemShut {NoStop}%
\bibitem [{\citenamefont {B{\o}ggild}\ \emph {et~al.}(2017)\citenamefont
  {B{\o}ggild}, \citenamefont {Caridad}, \citenamefont {Stampfer},
  \citenamefont {Calogero}, \citenamefont {Papior},\ and\ \citenamefont
  {Brandbyge}}]{Boggild2017}%
  \BibitemOpen
  \bibfield  {author} {\bibinfo {author} {\bibfnamefont {P.}~\bibnamefont
  {B{\o}ggild}}, \bibinfo {author} {\bibfnamefont {J.~M.}\ \bibnamefont
  {Caridad}}, \bibinfo {author} {\bibfnamefont {C.}~\bibnamefont {Stampfer}},
  \bibinfo {author} {\bibfnamefont {G.}~\bibnamefont {Calogero}}, \bibinfo
  {author} {\bibfnamefont {N.~R.}\ \bibnamefont {Papior}}, \ and\ \bibinfo
  {author} {\bibfnamefont {M.}~\bibnamefont {Brandbyge}},\ }\href {\doibase
  10.1038/ncomms15783} {\bibfield  {journal} {\bibinfo  {journal} {Nature
  Communications}\ }\textbf {\bibinfo {volume} {8}} (\bibinfo {year} {2017}),\
  10.1038/ncomms15783}\BibitemShut {NoStop}%
\bibitem [{\citenamefont {Brun}\ \emph {et~al.}(2019)\citenamefont {Brun},
  \citenamefont {Moreau}, \citenamefont {Somanchi}, \citenamefont {Nguyen},
  \citenamefont {Watanabe}, \citenamefont {Taniguchi}, \citenamefont
  {Charlier}, \citenamefont {Stampfer},\ and\ \citenamefont
  {Hackens}}]{Brun2019}%
  \BibitemOpen
  \bibfield  {author} {\bibinfo {author} {\bibfnamefont {B.}~\bibnamefont
  {Brun}}, \bibinfo {author} {\bibfnamefont {N.}~\bibnamefont {Moreau}},
  \bibinfo {author} {\bibfnamefont {S.}~\bibnamefont {Somanchi}}, \bibinfo
  {author} {\bibfnamefont {V.-H.}\ \bibnamefont {Nguyen}}, \bibinfo {author}
  {\bibfnamefont {K.}~\bibnamefont {Watanabe}}, \bibinfo {author}
  {\bibfnamefont {T.}~\bibnamefont {Taniguchi}}, \bibinfo {author}
  {\bibfnamefont {J.-C.}\ \bibnamefont {Charlier}}, \bibinfo {author}
  {\bibfnamefont {C.}~\bibnamefont {Stampfer}}, \ and\ \bibinfo {author}
  {\bibfnamefont {B.}~\bibnamefont {Hackens}},\ }\href {\doibase
  10.1103/physrevb.100.041401} {\bibfield  {journal} {\bibinfo  {journal}
  {Physical Review B}\ }\textbf {\bibinfo {volume} {100}} (\bibinfo {year}
  {2019}),\ 10.1103/physrevb.100.041401}\BibitemShut {NoStop}%
\bibitem [{\citenamefont {Pereira}\ \emph {et~al.}(2006)\citenamefont
  {Pereira}, \citenamefont {Mlinar}, \citenamefont {Peeters},\ and\
  \citenamefont {Vasilopoulos}}]{PMPV06}%
  \BibitemOpen
  \bibfield  {author} {\bibinfo {author} {\bibfnamefont {J.~M.}\ \bibnamefont
  {Pereira}}, \bibinfo {author} {\bibfnamefont {V.}~\bibnamefont {Mlinar}},
  \bibinfo {author} {\bibfnamefont {F.~M.}\ \bibnamefont {Peeters}}, \ and\
  \bibinfo {author} {\bibfnamefont {P.}~\bibnamefont {Vasilopoulos}},\ }\href
  {\doibase 10.1103/PhysRevB.74.045424} {\bibfield  {journal} {\bibinfo
  {journal} {Phys. Rev. B}\ }\textbf {\bibinfo {volume} {74}},\ \bibinfo
  {pages} {045424} (\bibinfo {year} {2006})}\BibitemShut {NoStop}%
\bibitem [{\citenamefont {Beenakker}\ \emph {et~al.}(2009)\citenamefont
  {Beenakker}, \citenamefont {Sepkhanov}, \citenamefont {Akhmerov},\ and\
  \citenamefont {Tworzyd{\l}o}}]{Beenakker2009}%
  \BibitemOpen
  \bibfield  {author} {\bibinfo {author} {\bibfnamefont {C.~W.~J.}\
  \bibnamefont {Beenakker}}, \bibinfo {author} {\bibfnamefont {R.~A.}\
  \bibnamefont {Sepkhanov}}, \bibinfo {author} {\bibfnamefont {A.~R.}\
  \bibnamefont {Akhmerov}}, \ and\ \bibinfo {author} {\bibfnamefont
  {J.}~\bibnamefont {Tworzyd{\l}o}},\ }\href {\doibase
  10.1103/physrevlett.102.146804} {\bibfield  {journal} {\bibinfo  {journal}
  {Physical Review Letters}\ }\textbf {\bibinfo {volume} {102}} (\bibinfo
  {year} {2009}),\ 10.1103/physrevlett.102.146804}\BibitemShut {NoStop}%
\bibitem [{\citenamefont {Zhang}\ \emph {et~al.}(2009)\citenamefont {Zhang},
  \citenamefont {He},\ and\ \citenamefont {Chen}}]{Zhang2009}%
  \BibitemOpen
  \bibfield  {author} {\bibinfo {author} {\bibfnamefont {F.-M.}\ \bibnamefont
  {Zhang}}, \bibinfo {author} {\bibfnamefont {Y.}~\bibnamefont {He}}, \ and\
  \bibinfo {author} {\bibfnamefont {X.}~\bibnamefont {Chen}},\ }\href {\doibase
  10.1063/1.3143614} {\bibfield  {journal} {\bibinfo  {journal} {Applied
  Physics Letters}\ }\textbf {\bibinfo {volume} {94}},\ \bibinfo {pages}
  {212105} (\bibinfo {year} {2009})}\BibitemShut {NoStop}%
\bibitem [{\citenamefont {Williams}\ \emph {et~al.}(2011)\citenamefont
  {Williams}, \citenamefont {Low}, \citenamefont {Lundstrom},\ and\
  \citenamefont {Marcus}}]{Williams2011}%
  \BibitemOpen
  \bibfield  {author} {\bibinfo {author} {\bibfnamefont {J.~R.}\ \bibnamefont
  {Williams}}, \bibinfo {author} {\bibfnamefont {T.}~\bibnamefont {Low}},
  \bibinfo {author} {\bibfnamefont {M.~S.}\ \bibnamefont {Lundstrom}}, \ and\
  \bibinfo {author} {\bibfnamefont {C.~M.}\ \bibnamefont {Marcus}},\ }\href
  {\doibase 10.1038/nnano.2011.3} {\bibfield  {journal} {\bibinfo  {journal}
  {Nature Nanotechnology}\ }\textbf {\bibinfo {volume} {6}},\ \bibinfo {pages}
  {222} (\bibinfo {year} {2011})}\BibitemShut {NoStop}%
\bibitem [{\citenamefont {Liu}\ \emph {et~al.}(2015)\citenamefont {Liu},
  \citenamefont {Rickhaus}, \citenamefont {Makk}, \citenamefont
  {T{\'{o}}v{\'{a}}ri}, \citenamefont {Maurand}, \citenamefont {Tkatschenko},
  \citenamefont {Weiss}, \citenamefont {Sch\"{o}nenberger},\ and\ \citenamefont
  {Richter}}]{Liu2015}%
  \BibitemOpen
  \bibfield  {author} {\bibinfo {author} {\bibfnamefont {M.-H.}\ \bibnamefont
  {Liu}}, \bibinfo {author} {\bibfnamefont {P.}~\bibnamefont {Rickhaus}},
  \bibinfo {author} {\bibfnamefont {P.}~\bibnamefont {Makk}}, \bibinfo {author}
  {\bibfnamefont {E.}~\bibnamefont {T{\'{o}}v{\'{a}}ri}}, \bibinfo {author}
  {\bibfnamefont {R.}~\bibnamefont {Maurand}}, \bibinfo {author} {\bibfnamefont
  {F.}~\bibnamefont {Tkatschenko}}, \bibinfo {author} {\bibfnamefont
  {M.}~\bibnamefont {Weiss}}, \bibinfo {author} {\bibfnamefont
  {C.}~\bibnamefont {Sch\"{o}nenberger}}, \ and\ \bibinfo {author}
  {\bibfnamefont {K.}~\bibnamefont {Richter}},\ }\href {\doibase
  10.1103/physrevlett.114.036601} {\bibfield  {journal} {\bibinfo  {journal}
  {Physical Review Letters}\ }\textbf {\bibinfo {volume} {114}} (\bibinfo
  {year} {2015}),\ 10.1103/physrevlett.114.036601}\BibitemShut {NoStop}%
\bibitem [{\citenamefont {Rickhaus}\ \emph {et~al.}(2015)\citenamefont
  {Rickhaus}, \citenamefont {Liu}, \citenamefont {Makk}, \citenamefont
  {Maurand}, \citenamefont {Hess}, \citenamefont {Zihlmann}, \citenamefont
  {Weiss}, \citenamefont {Richter},\ and\ \citenamefont
  {Sch\"{o}nenberger}}]{RLM+15}%
  \BibitemOpen
  \bibfield  {author} {\bibinfo {author} {\bibfnamefont {P.}~\bibnamefont
  {Rickhaus}}, \bibinfo {author} {\bibfnamefont {M.-H.}\ \bibnamefont {Liu}},
  \bibinfo {author} {\bibfnamefont {P.}~\bibnamefont {Makk}}, \bibinfo {author}
  {\bibfnamefont {R.}~\bibnamefont {Maurand}}, \bibinfo {author} {\bibfnamefont
  {S.}~\bibnamefont {Hess}}, \bibinfo {author} {\bibfnamefont {S.}~\bibnamefont
  {Zihlmann}}, \bibinfo {author} {\bibfnamefont {M.}~\bibnamefont {Weiss}},
  \bibinfo {author} {\bibfnamefont {K.}~\bibnamefont {Richter}}, \ and\
  \bibinfo {author} {\bibfnamefont {C.}~\bibnamefont {Sch\"{o}nenberger}},\
  }\href {https://epub.uni-regensburg.de/32427/} {\bibfield  {journal}
  {\bibinfo  {journal} {Nano Letters}\ }\textbf {\bibinfo {volume} {15}},\
  \bibinfo {pages} {5819} (\bibinfo {year} {2015})}\BibitemShut {NoStop}%
\bibitem [{\citenamefont {Cheng}\ \emph {et~al.}(2019)\citenamefont {Cheng},
  \citenamefont {Taniguchi}, \citenamefont {Watanabe}, \citenamefont {Kim},\
  and\ \citenamefont {Pillet}}]{Cheng2019}%
  \BibitemOpen
  \bibfield  {author} {\bibinfo {author} {\bibfnamefont {A.}~\bibnamefont
  {Cheng}}, \bibinfo {author} {\bibfnamefont {T.}~\bibnamefont {Taniguchi}},
  \bibinfo {author} {\bibfnamefont {K.}~\bibnamefont {Watanabe}}, \bibinfo
  {author} {\bibfnamefont {P.}~\bibnamefont {Kim}}, \ and\ \bibinfo {author}
  {\bibfnamefont {J.-D.}\ \bibnamefont {Pillet}},\ }\href {\doibase
  10.1103/physrevlett.123.216804} {\bibfield  {journal} {\bibinfo  {journal}
  {Physical Review Letters}\ }\textbf {\bibinfo {volume} {123}} (\bibinfo
  {year} {2019}),\ 10.1103/physrevlett.123.216804}\BibitemShut {NoStop}%
\bibitem [{\citenamefont {Zhao}\ \emph {et~al.}(2015)\citenamefont {Zhao},
  \citenamefont {Wyrick}, \citenamefont {Natterer}, \citenamefont
  {Rodriguez-Nieva}, \citenamefont {Lewandowski}, \citenamefont {Watanabe},
  \citenamefont {Taniguchi}, \citenamefont {Levitov}, \citenamefont
  {Zhitenev},\ and\ \citenamefont {Stroscio}}]{ZWN+15}%
  \BibitemOpen
  \bibfield  {author} {\bibinfo {author} {\bibfnamefont {Y.}~\bibnamefont
  {Zhao}}, \bibinfo {author} {\bibfnamefont {J.}~\bibnamefont {Wyrick}},
  \bibinfo {author} {\bibfnamefont {F.~D.}\ \bibnamefont {Natterer}}, \bibinfo
  {author} {\bibfnamefont {J.~F.}\ \bibnamefont {Rodriguez-Nieva}}, \bibinfo
  {author} {\bibfnamefont {C.}~\bibnamefont {Lewandowski}}, \bibinfo {author}
  {\bibfnamefont {K.}~\bibnamefont {Watanabe}}, \bibinfo {author}
  {\bibfnamefont {T.}~\bibnamefont {Taniguchi}}, \bibinfo {author}
  {\bibfnamefont {L.~S.}\ \bibnamefont {Levitov}}, \bibinfo {author}
  {\bibfnamefont {N.~B.}\ \bibnamefont {Zhitenev}}, \ and\ \bibinfo {author}
  {\bibfnamefont {J.~A.}\ \bibnamefont {Stroscio}},\ }\href {\doibase
  10.1126/science.aaa7469} {\bibfield  {journal} {\bibinfo  {journal}
  {Science}\ }\textbf {\bibinfo {volume} {348}},\ \bibinfo {pages} {672}
  (\bibinfo {year} {2015})}\BibitemShut {NoStop}%
\bibitem [{\citenamefont {Rodriguez-Nieva}\ and\ \citenamefont
  {Levitov}(2016)}]{RL15}%
  \BibitemOpen
  \bibfield  {author} {\bibinfo {author} {\bibfnamefont {J.~F.}\ \bibnamefont
  {Rodriguez-Nieva}}\ and\ \bibinfo {author} {\bibfnamefont {L.~S.}\
  \bibnamefont {Levitov}},\ }\href {\doibase 10.1103/physrevb.94.235406}
  {\bibfield  {journal} {\bibinfo  {journal} {Physical Review B}\ }\textbf
  {\bibinfo {volume} {94}} (\bibinfo {year} {2016}),\
  10.1103/physrevb.94.235406}\BibitemShut {NoStop}%
\bibitem [{\citenamefont {Bardarson}\ \emph {et~al.}(2009)\citenamefont
  {Bardarson}, \citenamefont {Titov},\ and\ \citenamefont {Brouwer}}]{BTB09}%
  \BibitemOpen
  \bibfield  {author} {\bibinfo {author} {\bibfnamefont {J.~H.}\ \bibnamefont
  {Bardarson}}, \bibinfo {author} {\bibfnamefont {M.}~\bibnamefont {Titov}}, \
  and\ \bibinfo {author} {\bibfnamefont {P.~W.}\ \bibnamefont {Brouwer}},\
  }\href {\doibase 10.1103/physrevlett.102.226803} {\bibfield  {journal}
  {\bibinfo  {journal} {Physical Review Letters}\ }\textbf {\bibinfo {volume}
  {102}} (\bibinfo {year} {2009}),\ 10.1103/physrevlett.102.226803}\BibitemShut
  {NoStop}%
\bibitem [{\citenamefont {Wurm}\ \emph {et~al.}(2011)\citenamefont {Wurm},
  \citenamefont {Richter},\ and\ \citenamefont {Adagideli}}]{WAR11}%
  \BibitemOpen
  \bibfield  {author} {\bibinfo {author} {\bibfnamefont {J.}~\bibnamefont
  {Wurm}}, \bibinfo {author} {\bibfnamefont {K.}~\bibnamefont {Richter}}, \
  and\ \bibinfo {author} {\bibfnamefont {I.}~\bibnamefont {Adagideli}},\ }\href
  {\doibase 10.1103/PhysRevB.84.075468} {\bibfield  {journal} {\bibinfo
  {journal} {Phys. Rev. B}\ }\textbf {\bibinfo {volume} {84}},\ \bibinfo
  {pages} {075468} (\bibinfo {year} {2011})}\BibitemShut {NoStop}%
\bibitem [{\citenamefont {N\"{o}ckel}\ and\ \citenamefont
  {Stone}(1997)}]{NS97}%
  \BibitemOpen
  \bibfield  {author} {\bibinfo {author} {\bibfnamefont {J.~U.}\ \bibnamefont
  {N\"{o}ckel}}\ and\ \bibinfo {author} {\bibfnamefont {A.~D.}\ \bibnamefont
  {Stone}},\ }\href {\doibase 10.1038/385045a0} {\bibfield  {journal} {\bibinfo
   {journal} {Nature}\ }\textbf {\bibinfo {volume} {385}},\ \bibinfo {pages}
  {45} (\bibinfo {year} {1997})}\BibitemShut {NoStop}%
\bibitem [{\citenamefont {Hentschel}\ and\ \citenamefont
  {Richter}(2002)}]{annbill}%
  \BibitemOpen
  \bibfield  {author} {\bibinfo {author} {\bibfnamefont {M.}~\bibnamefont
  {Hentschel}}\ and\ \bibinfo {author} {\bibfnamefont {K.}~\bibnamefont
  {Richter}},\ }\href {\doibase 10.1103/physreve.66.056207} {\bibfield
  {journal} {\bibinfo  {journal} {Physical Review E}\ }\textbf {\bibinfo
  {volume} {66}} (\bibinfo {year} {2002}),\
  10.1103/physreve.66.056207}\BibitemShut {NoStop}%
\bibitem [{\citenamefont {Wiersig}\ and\ \citenamefont
  {Hentschel}(2008)}]{reviewletters2008}%
  \BibitemOpen
  \bibfield  {author} {\bibinfo {author} {\bibfnamefont {J.}~\bibnamefont
  {Wiersig}}\ and\ \bibinfo {author} {\bibfnamefont {M.}~\bibnamefont
  {Hentschel}},\ }\href {\doibase 10.1103/PhysRevLett.100.033901} {\bibfield
  {journal} {\bibinfo  {journal} {Phys. Rev. Lett.}\ }\textbf {\bibinfo
  {volume} {100}},\ \bibinfo {pages} {033901} (\bibinfo {year}
  {2008})}\BibitemShut {NoStop}%
\bibitem [{\citenamefont {Schermer}\ \emph {et~al.}(2015)\citenamefont
  {Schermer}, \citenamefont {Bittner}, \citenamefont {Singh}, \citenamefont
  {Ulysse}, \citenamefont {Lebental},\ and\ \citenamefont
  {Wiersig}}]{applphyslett2015}%
  \BibitemOpen
  \bibfield  {author} {\bibinfo {author} {\bibfnamefont {M.}~\bibnamefont
  {Schermer}}, \bibinfo {author} {\bibfnamefont {S.}~\bibnamefont {Bittner}},
  \bibinfo {author} {\bibfnamefont {G.}~\bibnamefont {Singh}}, \bibinfo
  {author} {\bibfnamefont {C.}~\bibnamefont {Ulysse}}, \bibinfo {author}
  {\bibfnamefont {M.}~\bibnamefont {Lebental}}, \ and\ \bibinfo {author}
  {\bibfnamefont {J.}~\bibnamefont {Wiersig}},\ }\href {\doibase
  10.1063/1.4914498} {\bibfield  {journal} {\bibinfo  {journal} {Applied
  Physics Letters}\ }\textbf {\bibinfo {volume} {106}},\ \bibinfo {pages}
  {101107} (\bibinfo {year} {2015})},\ \Eprint
  {http://arxiv.org/abs/https://doi.org/10.1063/1.4914498}
  {https://doi.org/10.1063/1.4914498} \BibitemShut {NoStop}%
\bibitem [{\citenamefont {Xu}\ \emph {et~al.}(2018)\citenamefont {Xu},
  \citenamefont {Wang}, \citenamefont {Huang},\ and\ \citenamefont
  {Lai}}]{Lai18}%
  \BibitemOpen
  \bibfield  {author} {\bibinfo {author} {\bibfnamefont {H.-Y.}\ \bibnamefont
  {Xu}}, \bibinfo {author} {\bibfnamefont {G.-L.}\ \bibnamefont {Wang}},
  \bibinfo {author} {\bibfnamefont {L.}~\bibnamefont {Huang}}, \ and\ \bibinfo
  {author} {\bibfnamefont {Y.-C.}\ \bibnamefont {Lai}},\ }\href {\doibase
  10.1103/PhysRevLett.120.124101} {\bibfield  {journal} {\bibinfo  {journal}
  {Phys. Rev. Lett.}\ }\textbf {\bibinfo {volume} {120}},\ \bibinfo {pages}
  {124101} (\bibinfo {year} {2018})}\BibitemShut {NoStop}%
\bibitem [{\citenamefont {Han}\ \emph {et~al.}(2018)\citenamefont {Han},
  \citenamefont {Wang}, \citenamefont {Xu}, \citenamefont {Huang},\ and\
  \citenamefont {Lai}}]{Lai18a}%
  \BibitemOpen
  \bibfield  {author} {\bibinfo {author} {\bibfnamefont {C.-D.}\ \bibnamefont
  {Han}}, \bibinfo {author} {\bibfnamefont {C.-Z.}\ \bibnamefont {Wang}},
  \bibinfo {author} {\bibfnamefont {H.-Y.}\ \bibnamefont {Xu}}, \bibinfo
  {author} {\bibfnamefont {D.}~\bibnamefont {Huang}}, \ and\ \bibinfo {author}
  {\bibfnamefont {Y.-C.}\ \bibnamefont {Lai}},\ }\href {\doibase
  10.1103/physrevb.98.104308} {\bibfield  {journal} {\bibinfo  {journal}
  {Physical Review B}\ }\textbf {\bibinfo {volume} {98}} (\bibinfo {year}
  {2018}),\ 10.1103/physrevb.98.104308}\BibitemShut {NoStop}%
\bibitem [{\citenamefont {Handschin}\ \emph {et~al.}(2015)\citenamefont
  {Handschin}, \citenamefont {F\"{u}l\"{o}p}, \citenamefont {Makk},
  \citenamefont {Blanter}, \citenamefont {Weiss}, \citenamefont {Watanabe},
  \citenamefont {Taniguchi}, \citenamefont {Csonka},\ and\ \citenamefont
  {Sch\"{o}nenberger}}]{HFM+15}%
  \BibitemOpen
  \bibfield  {author} {\bibinfo {author} {\bibfnamefont {C.}~\bibnamefont
  {Handschin}}, \bibinfo {author} {\bibfnamefont {B.}~\bibnamefont
  {F\"{u}l\"{o}p}}, \bibinfo {author} {\bibfnamefont {P.}~\bibnamefont {Makk}},
  \bibinfo {author} {\bibfnamefont {S.}~\bibnamefont {Blanter}}, \bibinfo
  {author} {\bibfnamefont {M.}~\bibnamefont {Weiss}}, \bibinfo {author}
  {\bibfnamefont {K.}~\bibnamefont {Watanabe}}, \bibinfo {author}
  {\bibfnamefont {T.}~\bibnamefont {Taniguchi}}, \bibinfo {author}
  {\bibfnamefont {S.}~\bibnamefont {Csonka}}, \ and\ \bibinfo {author}
  {\bibfnamefont {C.}~\bibnamefont {Sch\"{o}nenberger}},\ }\href {\doibase
  10.1063/1.4935032} {\bibfield  {journal} {\bibinfo  {journal} {Applied
  Physics Letters}\ }\textbf {\bibinfo {volume} {107}},\ \bibinfo {pages}
  {183108} (\bibinfo {year} {2015})}\BibitemShut {NoStop}%
\bibitem [{\citenamefont {Wang}\ \emph {et~al.}(2009)\citenamefont {Wang},
  \citenamefont {Yan}, \citenamefont {Diehl}, \citenamefont {Hentschel},
  \citenamefont {Wiersig}, \citenamefont {Yu}, \citenamefont {Pfl\"{u}gl},
  \citenamefont {Belkin}, \citenamefont {Edamura}, \citenamefont {Yamanishi},
  \citenamefont {Kan},\ and\ \citenamefont {Capasso}}]{limacon_NJPhys}%
  \BibitemOpen
  \bibfield  {author} {\bibinfo {author} {\bibfnamefont {Q.~J.}\ \bibnamefont
  {Wang}}, \bibinfo {author} {\bibfnamefont {C.}~\bibnamefont {Yan}}, \bibinfo
  {author} {\bibfnamefont {L.}~\bibnamefont {Diehl}}, \bibinfo {author}
  {\bibfnamefont {M.}~\bibnamefont {Hentschel}}, \bibinfo {author}
  {\bibfnamefont {J.}~\bibnamefont {Wiersig}}, \bibinfo {author} {\bibfnamefont
  {N.}~\bibnamefont {Yu}}, \bibinfo {author} {\bibfnamefont {C.}~\bibnamefont
  {Pfl\"{u}gl}}, \bibinfo {author} {\bibfnamefont {M.~A.}\ \bibnamefont
  {Belkin}}, \bibinfo {author} {\bibfnamefont {T.}~\bibnamefont {Edamura}},
  \bibinfo {author} {\bibfnamefont {M.}~\bibnamefont {Yamanishi}}, \bibinfo
  {author} {\bibfnamefont {H.}~\bibnamefont {Kan}}, \ and\ \bibinfo {author}
  {\bibfnamefont {F.}~\bibnamefont {Capasso}},\ }\href {\doibase
  10.1088/1367-2630/11/12/125018} {\bibfield  {journal} {\bibinfo  {journal}
  {New Journal of Physics}\ }\textbf {\bibinfo {volume} {11}},\ \bibinfo
  {pages} {125018} (\bibinfo {year} {2009})}\BibitemShut {NoStop}%
\bibitem [{\citenamefont {Song}\ \emph {et~al.}(2009)\citenamefont {Song},
  \citenamefont {Fang}, \citenamefont {Liu}, \citenamefont {Ho}, \citenamefont
  {Solomon},\ and\ \citenamefont {Cao}}]{limacon_Cao2009}%
  \BibitemOpen
  \bibfield  {author} {\bibinfo {author} {\bibfnamefont {Q.}~\bibnamefont
  {Song}}, \bibinfo {author} {\bibfnamefont {W.}~\bibnamefont {Fang}}, \bibinfo
  {author} {\bibfnamefont {B.}~\bibnamefont {Liu}}, \bibinfo {author}
  {\bibfnamefont {S.-T.}\ \bibnamefont {Ho}}, \bibinfo {author} {\bibfnamefont
  {G.~S.}\ \bibnamefont {Solomon}}, \ and\ \bibinfo {author} {\bibfnamefont
  {H.}~\bibnamefont {Cao}},\ }\href {\doibase 10.1103/PhysRevA.80.041807}
  {\bibfield  {journal} {\bibinfo  {journal} {Phys. Rev. A}\ }\textbf {\bibinfo
  {volume} {80}},\ \bibinfo {pages} {041807} (\bibinfo {year}
  {2009})}\BibitemShut {NoStop}%
\bibitem [{\citenamefont {Yi}\ \emph {et~al.}(2009)\citenamefont {Yi},
  \citenamefont {Kim},\ and\ \citenamefont {Kim}}]{limacon_Kim2009}%
  \BibitemOpen
  \bibfield  {author} {\bibinfo {author} {\bibfnamefont {C.-H.}\ \bibnamefont
  {Yi}}, \bibinfo {author} {\bibfnamefont {M.-W.}\ \bibnamefont {Kim}}, \ and\
  \bibinfo {author} {\bibfnamefont {C.-M.}\ \bibnamefont {Kim}},\ }\href
  {\doibase 10.1063/1.3242014} {\bibfield  {journal} {\bibinfo  {journal}
  {Applied Physics Letters}\ }\textbf {\bibinfo {volume} {95}},\ \bibinfo
  {pages} {141107} (\bibinfo {year} {2009})}\BibitemShut {NoStop}%
\bibitem [{\citenamefont {Shinohara}\ \emph {et~al.}(2009)\citenamefont
  {Shinohara}, \citenamefont {Hentschel}, \citenamefont {Wiersig},
  \citenamefont {Sasaki},\ and\ \citenamefont
  {Harayama}}]{limacon_Susumu_Taka2009}%
  \BibitemOpen
  \bibfield  {author} {\bibinfo {author} {\bibfnamefont {S.}~\bibnamefont
  {Shinohara}}, \bibinfo {author} {\bibfnamefont {M.}~\bibnamefont
  {Hentschel}}, \bibinfo {author} {\bibfnamefont {J.}~\bibnamefont {Wiersig}},
  \bibinfo {author} {\bibfnamefont {T.}~\bibnamefont {Sasaki}}, \ and\ \bibinfo
  {author} {\bibfnamefont {T.}~\bibnamefont {Harayama}},\ }\href {\doibase
  10.1103/PhysRevA.80.031801} {\bibfield  {journal} {\bibinfo  {journal} {Phys.
  Rev. A}\ }\textbf {\bibinfo {volume} {80}},\ \bibinfo {pages} {031801}
  (\bibinfo {year} {2009})}\BibitemShut {NoStop}%
\bibitem [{\citenamefont {Klein}(1929)}]{Klein_ZPhys}%
  \BibitemOpen
  \bibfield  {author} {\bibinfo {author} {\bibfnamefont {O.}~\bibnamefont
  {Klein}},\ }\href {\doibase 10.1007/bf01339716} {\bibfield  {journal}
  {\bibinfo  {journal} {Zeitschrift f\"{u}r Physik}\ }\textbf {\bibinfo
  {volume} {53}},\ \bibinfo {pages} {157} (\bibinfo {year} {1929})}\BibitemShut
  {NoStop}%
\bibitem [{\citenamefont {Datta}(1995)}]{Datta1995}%
  \BibitemOpen
  \bibfield  {author} {\bibinfo {author} {\bibfnamefont {S.}~\bibnamefont
  {Datta}},\ }\href@noop {} {\emph {\bibinfo {title} {{Electronic Transport in
  Mesoscopic Systems}}}}\ (\bibinfo  {publisher} {Cambridge University Press},\
  \bibinfo {address} {Cambridge},\ \bibinfo {year} {1995})\BibitemShut
  {NoStop}%
\bibitem [{\citenamefont {Liu}\ \emph {et~al.}(2012)\citenamefont {Liu},
  \citenamefont {Bundesmann},\ and\ \citenamefont {Richter}}]{Liu2012}%
  \BibitemOpen
  \bibfield  {author} {\bibinfo {author} {\bibfnamefont {M.-H.}\ \bibnamefont
  {Liu}}, \bibinfo {author} {\bibfnamefont {J.}~\bibnamefont {Bundesmann}}, \
  and\ \bibinfo {author} {\bibfnamefont {K.}~\bibnamefont {Richter}},\ }\href
  {\doibase 10.1103/PhysRevB.85.085406} {\bibfield  {journal} {\bibinfo
  {journal} {Phys. Rev. B}\ }\textbf {\bibinfo {volume} {85}},\ \bibinfo
  {pages} {085406} (\bibinfo {year} {2012})}\BibitemShut {NoStop}%
\bibitem [{\citenamefont {Lee}\ \emph {et~al.}(2005)\citenamefont {Lee},
  \citenamefont {Ryu}, \citenamefont {Kwon}, \citenamefont {Rim},\ and\
  \citenamefont {Kim}}]{Lee2005}%
  \BibitemOpen
  \bibfield  {author} {\bibinfo {author} {\bibfnamefont {S.-Y.}\ \bibnamefont
  {Lee}}, \bibinfo {author} {\bibfnamefont {J.-W.}\ \bibnamefont {Ryu}},
  \bibinfo {author} {\bibfnamefont {T.-Y.}\ \bibnamefont {Kwon}}, \bibinfo
  {author} {\bibfnamefont {S.}~\bibnamefont {Rim}}, \ and\ \bibinfo {author}
  {\bibfnamefont {C.-M.}\ \bibnamefont {Kim}},\ }\href {\doibase
  10.1103/PhysRevA.72.061801} {\bibfield  {journal} {\bibinfo  {journal} {Phys.
  Rev. A}\ }\textbf {\bibinfo {volume} {72}},\ \bibinfo {pages} {061801}
  (\bibinfo {year} {2005})}\BibitemShut {NoStop}%
\bibitem [{\citenamefont {Groth}\ \emph {et~al.}(2014)\citenamefont {Groth},
  \citenamefont {Wimmer}, \citenamefont {Akhmerov},\ and\ \citenamefont
  {Waintal}}]{Groth2014}%
  \BibitemOpen
  \bibfield  {author} {\bibinfo {author} {\bibfnamefont {C.~W.}\ \bibnamefont
  {Groth}}, \bibinfo {author} {\bibfnamefont {M.}~\bibnamefont {Wimmer}},
  \bibinfo {author} {\bibfnamefont {A.~R.}\ \bibnamefont {Akhmerov}}, \ and\
  \bibinfo {author} {\bibfnamefont {X.}~\bibnamefont {Waintal}},\ }\href
  {http://stacks.iop.org/1367-2630/16/i=6/a=063065} {\bibfield  {journal}
  {\bibinfo  {journal} {New Journal of Physics}\ }\textbf {\bibinfo {volume}
  {16}},\ \bibinfo {pages} {063065} (\bibinfo {year} {2014})}\BibitemShut
  {NoStop}%
\bibitem [{\citenamefont {Hecht}(2002)}]{hechtoptics}%
  \BibitemOpen
  \bibfield  {author} {\bibinfo {author} {\bibfnamefont {E.}~\bibnamefont
  {Hecht}},\ }\href@noop {} {\emph {\bibinfo {title} {Optics}}}\ (\bibinfo
  {publisher} {Addison-Wesley},\ \bibinfo {year} {2002})\BibitemShut {NoStop}%
\bibitem [{\citenamefont {McCann}\ and\ \citenamefont
  {Koshino}(2013)}]{McCannReview2013}%
  \BibitemOpen
  \bibfield  {author} {\bibinfo {author} {\bibfnamefont {E.}~\bibnamefont
  {McCann}}\ and\ \bibinfo {author} {\bibfnamefont {M.}~\bibnamefont
  {Koshino}},\ }\href {\doibase 10.1088/0034-4885/76/5/056503} {\bibfield
  {journal} {\bibinfo  {journal} {Reports on Progress in Physics}\ }\textbf
  {\bibinfo {volume} {76}},\ \bibinfo {pages} {056503} (\bibinfo {year}
  {2013})}\BibitemShut {NoStop}%
\bibitem [{\citenamefont {St\"{o}ckmann}(1999)}]{stoecki_quantumchaos}%
  \BibitemOpen
  \bibfield  {author} {\bibinfo {author} {\bibfnamefont {H.-J.}\ \bibnamefont
  {St\"{o}ckmann}},\ }\href {\doibase 10.1017/CBO9780511524622} {\emph
  {\bibinfo {title} {Quantum Chaos: An Introduction}}}\ (\bibinfo  {publisher}
  {Cambridge University Press},\ \bibinfo {year} {1999})\BibitemShut {NoStop}%
\end{thebibliography}%
%
%
%
%

\end{document}